\documentclass[english,aps,preprint]{revtex4}
\usepackage[latin1]{inputenc}
                                                                                     
\makeatletter
                                                                                     
\usepackage[dvips]{graphicx}% Include figure files
\usepackage{dcolumn}% Align table columns on decimal point
\usepackage{bm}% bold math                                                                                     
\usepackage{babel}
\usepackage{longtable}
\begin{document}
                                                                                     
\preprint{303431JCP $Revision: 2.2 $}
%JM note a bunch of hacks were necessary to make the file
%JM "play well" with the PDF hyperlinking of arXiv.org
\title{Development of Novel Density Functionals for Thermochemical Kinetics}
                                                                                     
\author{A. Daniel Boese}
\email{daniel.boese@weizmann.ac.il}
\author{Jan M. L. Martin}
\email{comartin@wicc.weizmann.ac.il}
                                                                                     
\affiliation{Department of Organic Chemistry, Weizmann Institute of Science, IL-76100 Re\d{h}ovot, Israel}
\date{{\em J. Chem. Phys.}, in press ({\bf 303431JCP}); Received April 9, 2004; Accepted May 28, 2004}

\begin{abstract}
A new density functional theory (DFT) exchange-correlation functional for the exploration of reaction mechanisms is proposed.
This new functional, denoted BMK (Boese-Martin for Kinetics),
has an accuracy in the 2 kcal/mol range for
transition state barriers
but, unlike previous attempts at such a
functional, this improved accuracy does {\em not} come at the expense of equilibrium properties.
This makes it a general-purpose functional whose domain of applicability has been extended to
transition states, rather than a specialized functional for kinetics.
The improvement in BMK rests on the inclusion of the kinetic energy density together with a large value of
the exact exchange mixing coefficient. For this functional, the kinetic energy density appears to correct
`back' the excess exact exchange mixing for ground-state properties, possibly simulating variable
exchange.
\end{abstract}
\maketitle
                                                                                     
\section{introduction}

In the last decade, density functional theory (DFT) has probably become the number one applied computational chemistry tool.
Most exchange-correlation functionals currently used\cite{VWN,PW91,B88X,P86,P91c,LYP} --- including the exceedingly
popular Becke 3-parameter-Lee-Yang-Parr (B3LYP) hybrid functional \cite{B3P91,LYP} --- have been developed more than ten years ago.
Yet these first-generation functionals exhibit a number of chemically important weaknesses.

A number of second-generation functionals have been proposed as successors in recent years,
such as PBE (Perdew-Burke-Ernzerhof\cite{PBE}),
mPW91 (modified Perdew-Wang 1991\cite{mPW91}), 
VSXC (Van Voorhis-Scuseria exchange-correlation\cite{VSXC}),
PBE0 (PBE with 25\% exact exchange added\cite{PBE0}), 
PKZB (Perdew-Kurth-Zupan-Blaha\cite{PKZB}), 
B97 (Becke 1997\cite{B97}),
B97-1 (Becke 1997 variant 1\cite{HCTH}), 
B97-2 (Becke 1997 variant 2\cite{B972}),
the HCTH (Hamprecht-Cohen-Tozer-Handy) family\cite{HCTH,HCTH2,HCTH3},
OPTX/OPTC (optimized exchange/correlation\cite{OPTX,OPTC}),
$\tau$-HCTH and its hybrid\cite{tHCTH},
TPSS (Tao-Perdew-Staroverov-Scuseria\cite{TPSS}),
and B98 (Becke 1998 \cite{B98}).
Two important avenues that have been pursued in this direction are: (a) inclusion of new variables explicitly dependent on the Kohn-Sham orbitals, such as the
kinetic energy density $\tau$; (b) treatment of DFT as a parametrized `semi-ab initio' method\cite{EDF1}.

In particular, the most commonly used hybrid functionals all involve parametrization against experimental
data to some extent: For instance, the mixing parameters in B3LYP were fitted against the binding energies
of the molecules in the G2-1 set\cite{G2}. Becke's subsequent 1997 functional\cite{B97} involves a power 
series expansion in the reduced density gradient, parameters of which were fitted against the binding energies,
ionization potentials, and proton affinities in the G2-1 set
plus some additional data. An expansion of order $m$
requires $3m+4$ parameters: for $m=4$ and higher the fit became unphysical, and even $m=3$ appeared to exhibit
signs of over-fitting. 
Erring on the side of caution, Becke chose $m=2$, involving 10 parameters.

The issue of over-fitting may be substantially alleviated by (a) admitting high-level ab initio data to the `training set';
(b) considering data other than energetics, such as gradients at the equilibrium geometry (a convenient metric for
geometry errors) and numerical exchange-correlation potentials extracted from high-level ab initio electron densities
by means of the Zhao-Morrison-Parr\cite{ZMP} procedure.
Tozer and Handy\cite{TH1,TH2,TH3} followed this route, with only moderate success, as their functional
form was `poorly chosen'\cite{TH3}. Greater success was achieved by applying this more elaborate refinement procedure
to Becke's functional form, yielding the original HCTH/93 (Hamprecht-Cohen-Tozer-Handy\cite{HCTH,CH}) functional. 
Besides this $m=4$ `pure DFT' functional, a reparametrization of Becke's hybrid functional (denoted B97-1) was
obtained as a by-product.

The general performance of HCTH/93 was markedly superior to first-generation GGA (generalized gradient approximation) 
functionals such as BLYP (Becke-Lee-Yang-Parr\cite{B88X,LYP}). However, it did not measure up to expectations for weakly bound systems\cite{CHcomment,TBH}; in general, the absence of anions in the training set meant
that
the functional was not adequately
parametrized for the outer regions of the electron density. In an attempt to remedy this problem and generally increase
the versatility of the functional, it was reparametrized\cite{HCTH2,HCTH3} 
to ever larger sets of data including anions and weakly
bound systems, culminating in the HCTH/407 functional\cite{HCTH3}.
The numerals in the notation refer to the number of chemical systems used in the training set.

Attempts to include additional variables such as the kinetic energy density $\tau$ into the functional
have been reviewed in the introduction to Ref.\cite{tHCTH}.
This latter paper
represents
an attempt to 
enhance HCTH with $\tau$-dependent terms. One intriguing question here is the actual role played by the kinetic energy density.
Becke has argued\cite{Becke00} that local density functionals that include the kinetic energy
density $\tau$ can simulate delocalized exchange and thus, inclusion of $\tau$ will improve 
generalized gradient
functionals. Van Voorhis and Scuseria\cite{VSXC} presented evidence that such
a form might be capable of simulating exact exchange. Boese and Handy\cite{tHCTH} have similar 
experience, but conjectured that $\tau$ does not merely simulate exact exchange,
but rather {\em variable} exact exchange\cite{tHCTH}.

This brings us to the subject of this paper. One well-known deficiency (see e.g.\cite{baker95,durant96})
of most existing exchange-correlation functionals is their poor performance for reaction barrier heights:
These are often significantly underestimated, and it is indeed not unusual for DFT not to find a barrier
at all. Durant\cite{durant96} noted that BH\&HLYP (Becke half-and-half exchange with LYP correlation\cite{BHLYP}),
a hybrid functional with 50\% exact exchange, was the exception to the rule in that it predicted transition state
geometries and barriers reasonably well, despite being inferior to B3LYP (20\% exact exchange) for thermochemical
properties. Later, Truhlar and coworkers\cite{mPW1K} and Kang and Musgrave\cite{KMLYP} both proposed 
reparametrizations of existing functionals (modified Perdew-Wang\cite{mPW91} in the former,
B3LYP without Becke exchange in the latter case) for barrier heights: both functionals involve a very high fraction of exact exchange,
42.7\% in the case of mPW1K (modified Perdew-Wang for kinetics\cite{mPW1K}), 55.7\% in the
case of KMLYP (Kang-Musgrave-Lee-Yang-Parr\cite{KMLYP}). For both functionals, the
improved performance for reaction barrier heights comes at the expense of seriously degraded performance 
for ground-state properties: in particular, 
errors in atomization energies and geometries exceed those for more conventional
hybrid functionals
with exact exchange in the 15--25\% range by a factor of two to three\cite{basissetpaper}.
In the case of KMLYP, an empirical G2-style `high level correction' was proposed, but this is strictly a stop-gap solution
as it will not affect properties other than equilibrium energetics.

For researchers involved in the computational study of reaction mechanisms (such as our own group, see e.g.\cite{acetone,revital}) the 
present situation is very frustrating --- especially when a plethora of competing multi-step pathways
is to be considered\cite{acetone} --- as no single functional can be relied upon to get both relative stabilities and barrier heights
right. 

One possible route to such a functional might be variable exact exchange. If the kinetic energy density were truly capable of
simulating variable exact exchange, however, then a fixed-exact exchange functional with $\tau$-dependent terms might afford
a computationally much simpler solution. In the present paper, we will show that this is indeed the case, and shall propose
a new general-purpose exchange-correlation functional to be known as BMK (Boese-Martin for kinetics).

Before discussing the development of the new functional, however, we would like to mention a couple of
alternative
approaches to tackling
the aforementioned problem of transition states. These are connected to attempts by various groups at creating
`third-generation functionals'. None of these functionals are currently accurate enough to replace any of the second-generation
functionals.

Self-interaction-free
or -corrected functionals constitute the first possibility. 
Here, an additional orbital-dependent
correction is introduced\cite{SIDFT1}, causing the energy to lose invariance to rotations of occupied orbitals. Solving
these equations self-consistently is both time-consuming and difficult\cite{SIDFT2}. Furthermore, no gradients or higher derivatives with respect
to the geometry are available, which greatly reduces their usefulness in practical applications.

The second novel approach is to use functionals with so-called local exact exchange. However, such functionals either perform much
less well when describing ground-state properties than the known functionals\cite{localx1}, or have various difficulties
describing spin splittings\cite{localx2}.

\section{Computational Details}

In order to obtain additional reference data beyond those in the HCTH/407 data set\cite{HCTH3}, 
the W1 and W2 (Weizmann-1 and -2) computational thermochemistry protocols\cite{W2,W2validate} were applied. Basically
these methods represent approximations to the CCSD(T) infinite basis set limit: They generate average errors for
total atomization energies (TAEs) in the kJ/mol range, and should therefore definitely be accurate enough for our
purpose. The calculations were carried out by means of the MOLPRO2002 electronic structure package\cite{MOLPRO}
and a driver script written in Perl\cite{Onur}.

The parametrization of the functional itself was carried out using a collection of ad hoc programs and scripts written by one of us (ADB):
modified versions of both CADPAC\cite{Cadpac} and Gaussian 03\cite{g03} were used as DFT engines. 

Both are finite basis set DFT codes. The HCTH family of functionals was parametrized using a triple-zeta plus double polarization (TZ2P) 
basis set. We discovered early on in the present study that this basis set
is
insufficiently saturated for our purposes: we indeed found that 
in the case of the HCTH and $\tau$-HCTH functionals, some basis set superposition error has been absorbed into the
functional.
On the dependence of DFT parametrization on the basis set used, see Ref.\cite{basissetpaper}. After some 
convergence studies, we decided on the following combination of basis sets: Jensen aug-pc2 (augmented polarization consistent-2)
basis set\cite{Jensen} for \{H,C,N,O,F\}, TZ3Pf+diffuse (triple-zeta plus triple polarization plus diffuse, including 
an additional high-exponent $d$ function
as known to be required for many second-row compounds\cite{so2}) for \{B, Al, Si, P, S, Cl\}, the Wachters-Hay basis set\cite{Wachters}
for first-row transition metals augmented with $2f1g$ polarization functions taken from the Appendix to Ref.\cite{sdb-cc},
the SDB-aug-cc-pVTZ basis set\cite{sdb-cc} for third- and fourth-row main group elements, and the 
6-311+G(3df) basis set\cite{6-311dreck} for remaining elements.
We are confident that this combination of basis sets is sufficiently close to the Kohn-Sham basis set limit
for our purposes.

\section{Towards a New Functional}

We shall try to develop a functional with the following design goals:
\begin{itemize}
\item Mean absolute error below 2 kcal/mol for reaction barrier heights;
\item Smallest mean absolute error possible for atomization energies, geometry and 
harmonic frequencies
of ground-state molecules;
\item No separation between exchange and correlation functionals, in light of error cancellation between the two terms;
\item A
mathematically
`simple' functional in order to facilitate insight; %maximise the understanding from the effects captured
\item A power series expansion, yielding the maximum number of adjustable parameters %linear coefficients which can be determined 
(usually on the order of 10-20);
\item A functional that obeys some of the more simple scaling conditions and furthermore is capable of modeling the exchange hole\cite{B83,B98};
\item As diverse and balanced a parametrization set as possible, including transition metal (TM) complexes  and hydrogen-bonded systems.
%Careful determination of the functional regarding more weakly bound systems such as hydrogen bonds
\end{itemize}
We will consider two different functional forms, one the hybrid HCTH\cite{HCTH} form (of which B97\cite{B97} and B97-1\cite{HCTH} are special cases), the other the hybrid $\tau$-HCTH form.
The exchange-correlation energy (with a as the mixing coefficient) is defined as: %determined by:
\begin{eqnarray}
E_{XC}&=&E_{X,l}+E_{X,n-l}+E_C+a\times E_{HF}\\
E_{X,l}&=&\sum_{\sigma}\int e_{X\sigma}^{LSDA}(\rho_{\sigma})g_{X\sigma,l}(s_{\sigma}^2)d{\bf r}\\
g_{X\sigma,l}&=&\sum_{i=0}^Mc_{X\sigma,l,i}u_{X\sigma}^i\\
E_{X,n-l}&=&\sum_{\sigma}\int e_{X\sigma}^{LSDA}(\rho_{\sigma})g_{X\sigma,n-l}(s_{\sigma}^2)f_{X \sigma}(w_{\sigma})d{\bf r}\\
f_{X \sigma}(w_{\sigma})&=&w_{\sigma}-2(w_{\sigma})^3+(w_{\sigma})^5\\
w_{\sigma}&=&\frac{\frac{\frac{3}{5}(6\pi^2)^{2/3}\rho_{\sigma}^{5/3}}{\tau_{\sigma}}-1}
{\frac{\frac{3}{5}(6\pi^2)^{2/3}\rho_{\sigma}^{5/3}}{\tau_{\sigma}}+1}\\
g_{X\sigma,n-l}&=&\sum_{i=0}^Mc_{X\sigma,n-l,i}u_{X\sigma}^i\\
u_{X\sigma}&=&\gamma_{X\sigma}s_{\sigma}^2(1+\gamma_{X\sigma}s_{\sigma}^2)^{-1}\\
E_C&=&\sum_{\sigma}E_{C\sigma\sigma}+E_{C\alpha\beta}\\
E_{C\sigma\sigma}&=&\int e_{C\sigma\sigma}^{LSDA}(\rho_{\sigma})g_{C\sigma\sigma}(s_{\sigma}^2)d{\bf r}\\
g_{C\sigma\sigma}&=&\sum_{i=0}^Mc_{C\sigma\sigma,i}u_{C\sigma\sigma}^i\\
u_{C\sigma\sigma}&=&\gamma_{C\sigma\sigma}s_{\sigma}^2(1+\gamma_{C\sigma\sigma}s_{\sigma}^2)^{-1}\\
E_{C\alpha\beta}&=&\int e_{C\alpha\beta}(\rho_{\alpha},\rho_{\beta})g_{C\alpha\beta}(s_{avg}^2)d{\bf r}\\
g_{C\alpha\beta}&=&\sum_{i=0}^Mc_{C\alpha\beta,i}u_{C\alpha\beta}^i\\
u_{C\alpha\beta}&=&\gamma_{C\alpha\beta}s_{avg}^2(1+\gamma_{C\alpha\beta}s_{avg}^2)^{-1}
\end{eqnarray}
Here,
$s$ is closely related to the reduced gradient:
\begin{eqnarray}
s_{\sigma}^2&=&|\nabla\rho_{\sigma}|^2\rho_{\sigma}^{-8/3}\\
s_{avg}^2&=&\frac{1}{2}(s_{\alpha}^2+s_{\beta}^2)
\end{eqnarray}
The coefficients $\gamma$ are non-linear and fixed to:
\begin{eqnarray}
\gamma_{X\sigma}&=&0.004\\
\gamma_{C\sigma\sigma}&=&0.2\\
\gamma_{C\alpha\beta}&=&0.006\\
\end{eqnarray}
and the kinetic energy density $\tau$ is defined as:
\begin{eqnarray}
\tau_{\sigma}=\sum(\nabla\phi_{i\sigma})^2
\end{eqnarray}
For the B97 form, $E_{X,n-l}$ will be zero. The B97-1 functional\cite{HCTH} has 12 linear coefficients with the sum over M in equations 3, 11, and 13 going to M=2.
The $\tau$-HCTH hybrid functional\cite{tHCTH} had 16 linear coefficients with the sum over M in equations 3, 7, 11, and 13 going to M=3.
In both cases, the sum was truncated when no appreciable further improvement was seen from the inclusion of higher-order terms.
The main improvement in the $\tau$-HCTH hybrid functional obviously came from the higher order terms which included the kinetic energy density.

\begin{figure}
\includegraphics[height=12cm,width=18cm]{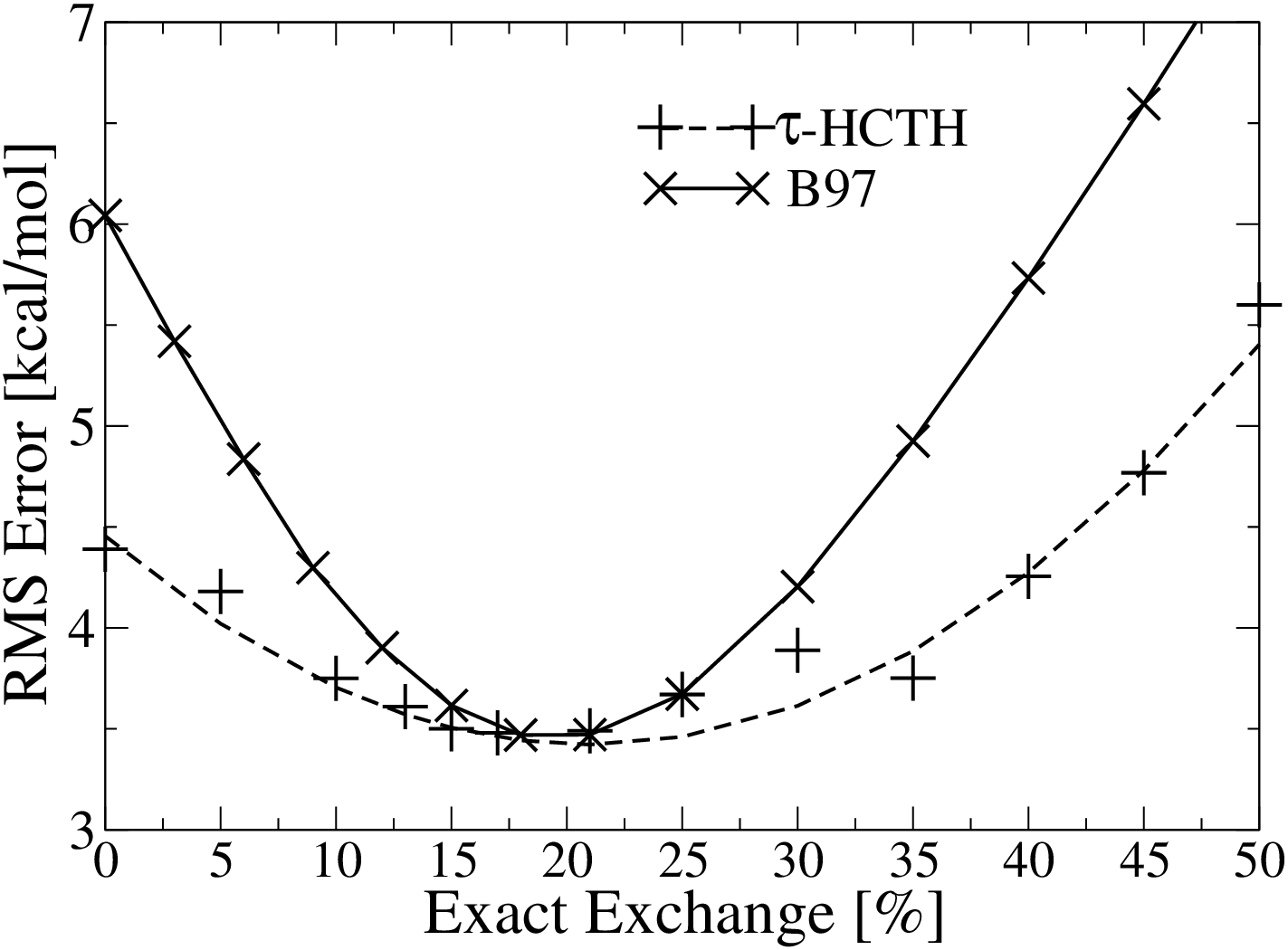}
\vspace{5cm}
\caption{\label{Fig1}Boese and Martin, Journal of Chemical Physics}
\end{figure}

In a first `numerical experiment', we varied the amount of Hartree-Fock exchange
(the variable a in equation 1)
in each of these functionals in small increments between 0 and 50~\%, and `self-consistently'
{\em re-refined all other parameters} to the HCTH/147 dataset\cite{HCTH3}
which is similar to the G2-2 dataset of Pople and coworkers\cite{G2,G22n,G22i}. The RMS errors for both `meta-functionals', as a 
function of the percentage of Hartree-Fock exchange, are plotted in Figure \ref{Fig1}.

We can make the following observations. First of all, both curves exhibit minima in the 20\% range,
consistent with experience. Second, the improvement afforded by the kinetic energy density terms is not nearly as large as one might have expected. Third, the main difference occurs far away from the minimum: the `meta-functional' including $\tau$ deteriorates much less rapidly than the one without, both as exact exchange is reduced to zero and as it is cranked up into the 50\% region.

It may be instructive to consider the evolution of the parameters --- particularly the zeroth-order
linear coefficients of equations 3--14 which contribute most to the energy --- as a function of 
`exact exchange' admixture.
All these coefficients, save the second non-local one, will be unity for the uniform electron gas.
For the B97-like form, the zeroth-order exchange-coefficient $c_{X\sigma,l,0}$ varies  from 1.1 (at 0\% HF exchange) to about 0.4
(at 50\% HF exchange). 
For the $\tau$-HCTH-like form, at 50\% exact exchange mixing, this coefficient stands at 0.5, or 1.2 higher
than without the terms including the kinetic energy density. As we would have expected, the $c_{X\sigma,n-l,0}$ coefficient
changes from near-zero (0.001) at 0\% HF exchange to -0.38 at 50\% HF exchange. Hence, the terms including the kinetic
energy become much larger, apparently `correcting back' the exact exchange mixing. 
Furthermore, the zeroth-order correlation
coefficients are affected a lot by adding exact exchange: For example, $c_{C\sigma\sigma,0}$ actually becomes negative. For
the B97 form, these coefficients change from 0.4 to -2.8. The $\tau$-HCTH form, however, behaves completely differently, with 
$c_{X\sigma\sigma,0}$ remaining around 0.4. For the B97 form, this effect is cancelled by the $c_{X\alpha\beta,0}$
coefficient, going from 0.5 to 1.8. Again, the $\tau$-HCTH form functionals coefficient does barely change from 0.7 to 0.9.
As a conclusion, we note that the terms depending on the kinetic energy density have to be responsible
for the relatively shallow `$\tau$-including' curve in Figure \ref{Fig1}. 
Moreover, changes in the coefficients are by and large limited to the exchange part, while for the $\tau$-free (B97) form,
both exchange and correlation coefficients are quite sensitive to variation in the amount of Hartree-Fock exchange.
Thus, in short, introduction of Equations 4 and 5 
into the functional makes it capable of correcting {\em back} for excess exact exchange.

This immediately begs the question whether this property of the $\tau$-HCTH hybrid form 
also applies to reaction barrier heights and transition state geometries. Furthermore, what would
be the behavior of the various terms and parameters as a function of the percentage of 
Hartree-Fock exchange?

For this purpose, we included such transition states into our `training' or `fitting' set. At this stage,
we employed the set of reactions proposed by Truhlar and coworkers\cite{mPW1Keval} for which 
accurate experimentally derived barrier heights (including tunneling and other quantum corrections)
are available. To this we added two identity S$_N$2 reactions for which accurate barriers were previous determined by W2 theory\cite{ParthiSN2}. For Truhlar's reactions, we performed W1 and W2 reference
calculations at the QCISD/MG3 geometries given in the Supporting Information to Ref.\cite{mPW1Keval}.
Results are summarized in Table \ref{tab1}.
Overall, the agreement between theory and the values derived from experiment is quite good: any discrepancies would be primarily due to post-CCSD(T) correlation effects, or uncertainty in the
measurements or quantum corrections.
The transition states of the reactions H$_2$ + F and OH + Cl exhibit severe nondynamical correlation, and CCSD(T) data would not be very meaningful. In the case of  the H$_2$ + F reaction, we prefer
to use the barrier of 2.31 kcal/mol obtained by K\'allay et al.\cite{KallayHF} in the set, and in the case of the OH + Cl reaction, we ended up employing the experimental
value of 9.8 kcal/mol. 
Altogether 44 barriers (22 forward and 22 reverse) are reported in  Table \ref{tab1}, although 
obviously, a method that would get both the forward barrier and the reaction energy right would
automatically give the right answer for the reverse barrier --- so either the reverse barrier height
or the reaction energy would be redundant data.
For determining the functional, we included
these 22 forward reaction barrier heights 
plus the reaction barriers of Cl + CH$_3$Cl $\rightarrow$ Cl + CH$_3$Cl and
F + CH$_3$F $\rightarrow$ F + CH$_3$F. Since only some of the molecules in Table \ref{tab1} were included in the HCTH/147 set, we
only use 14 of the 19 reaction energies in the small set. These were augmented with nine reaction energies\cite{Pflueger} that were used
by Sch\"utz et al. to assess local correlation methods\cite{LCCSD}.
 
If we were to exclusively consider errors in atomization energies, it is at least
in principle possible that just a particular type of bond is poorly described, and the
error statistics obtained are not representative for reactions involving formation
and/or breaking of other bond types. Adding reaction
energies to the training set
at least allows us to verify that the various molecules involved at least
yield correct energetics relative to their molecular `next of kin'. Furthermore,
typical DFT studies are probably more likely to focus on reaction energies than 
on atomization energies.

The original HCTH/147 set\cite{HCTH2} was thus enlarged by 22 reaction barrier heights, 23 reaction energies, the dissociation energy
of the NH$_3$ and (H$_2$O)(NH$_3$) dimers, and some potential energy points of the NH$_3$ dimer surface\cite{HCTH4}. The
latter two properties were given low weights, while the reaction energies were assigned 
ten times the weight
for the atomization energies,
thus taking into account that said reaction energies are an order of magnitude smaller than typical
atomization energies
multiplied by the molecularity (number of molecules involved) of the reaction.

\begin{figure}
\includegraphics[height=12cm,width=18cm]{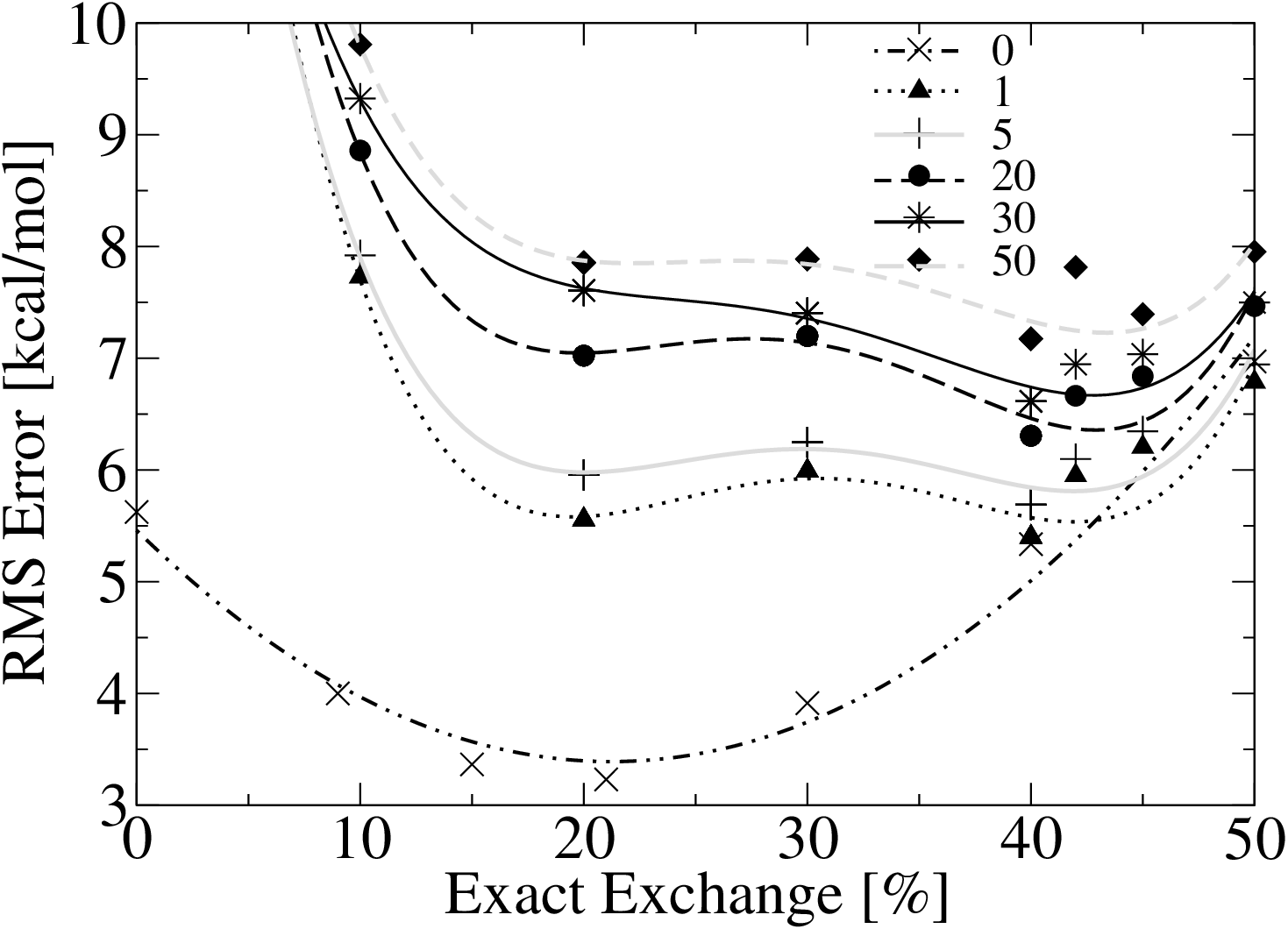}
\vspace{5cm}
\caption{\label{Fig2}Boese and Martin, Journal of Chemical Physics}
\end{figure}

\begin{figure}
\includegraphics[height=12cm,width=18cm]{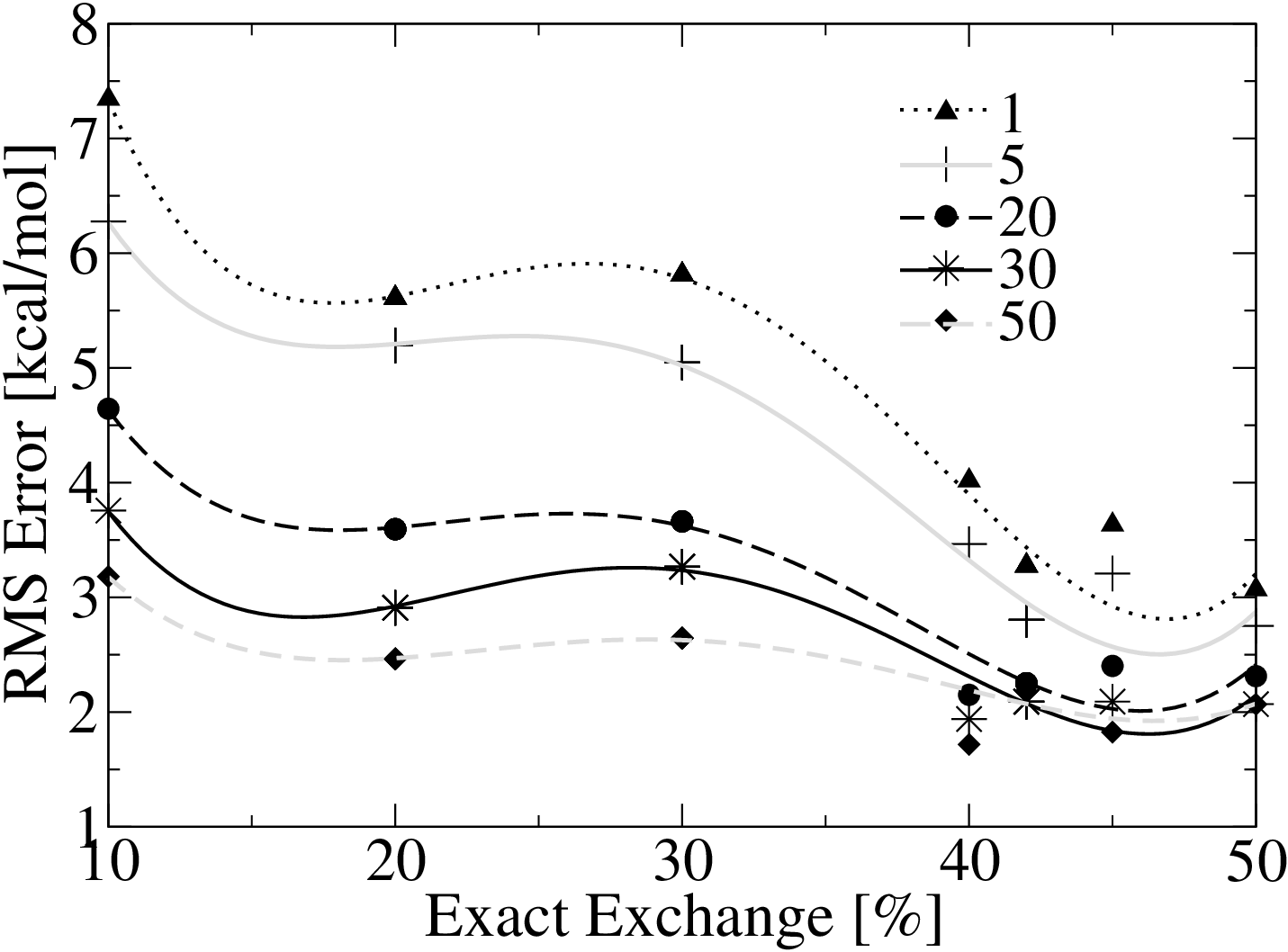}
\vspace{5cm}
\caption{\label{Fig3}Boese and Martin, Journal of Chemical Physics}
\end{figure}

Once again, all parameters were `self-consistently' refitted for every value of the exact exchange coefficient, and this
analysis was repeated for various relative weights assigned to the barrier heights in the `penalty function'.
For the $\tau$-free form, profiles of the
RMS errors for all energetics other than barrier heights are plotted in Figure \ref{Fig2} (every curve representing a different
weight factor for the barrier heights), while the corresponding profiles for the barrier heights themselves can be found
in Figure \ref{Fig3}. With a zero weight for the transition states, the expected single minimum around 20\% Hartree-Fock
exchange is obtained. However, as soon as the barriers are even given a unit weight, a second minimum appears
around 40\%. As the weight for the transition states is increased, both minima persist at least through $W_{\rm TS}=50$.
Note that the curves are not perfectly smooth since the final parameters are somewhat dependent on the initial 
parametrization chosen.

\begin{figure}
\includegraphics[height=12cm,width=18cm]{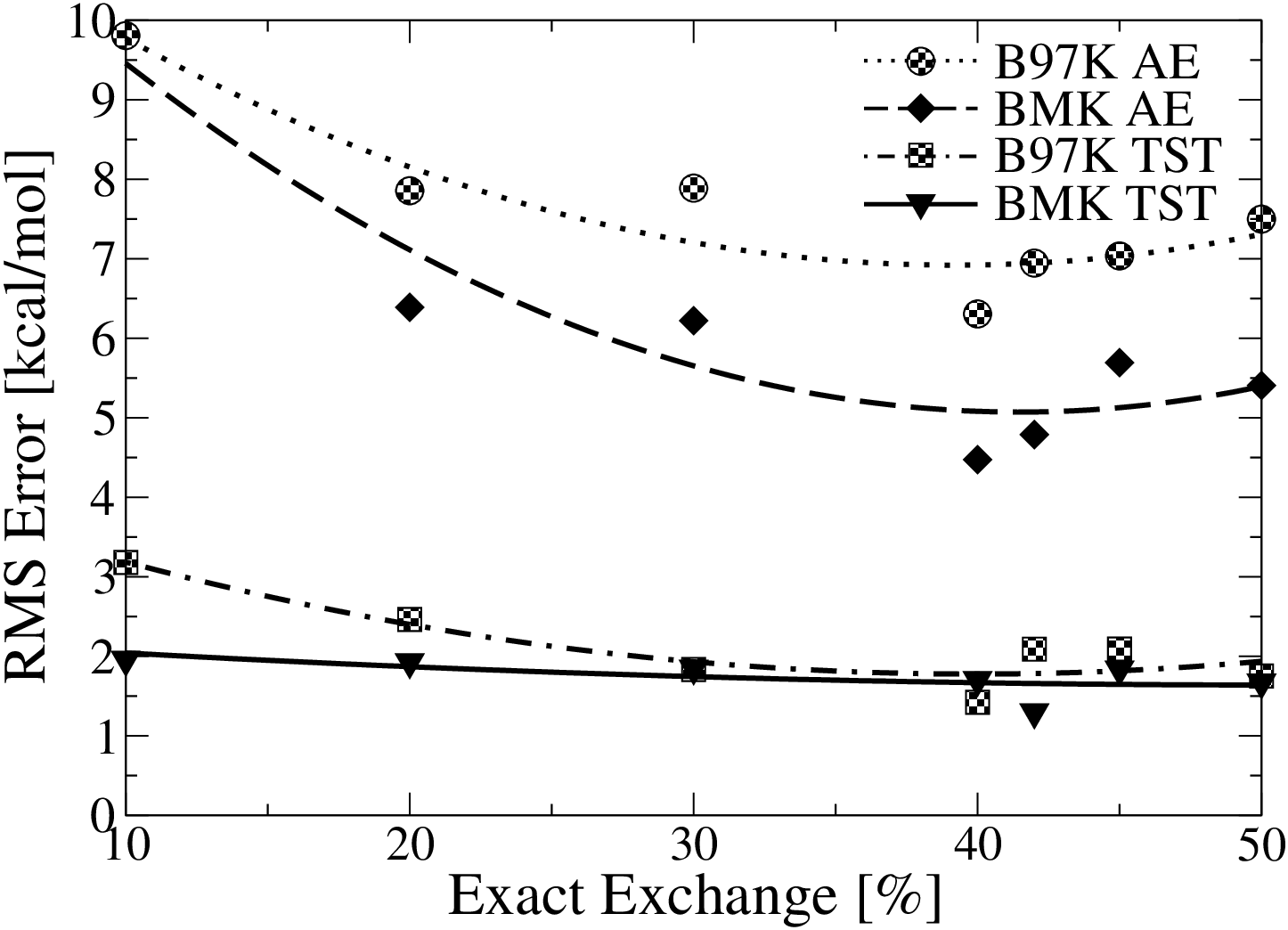}
\vspace{5cm}
\caption{\label{Fig4}Boese and Martin, Journal of Chemical Physics}
\end{figure}

For a functional to be useful for transition states, barriers have to be predicted quite accurately --- especially for reactions
with low barriers which might otherwise be spuriously rendered barrierless. A 2 kcal/mol RMS target accuracy would, 
according to Figure \ref{Fig3}, require $W_{\rm TS}=50$. Using this relative weight, we have plotted RMS error profiles
for both barrier heights and other energetic properties as a function of the percentage of Hartree-Fock exchange
in Figure \ref{Fig4}, for both the $\tau$-free and the $\tau$-including forms.
Parameter optimizations for the latter took the final parameters for the former as initial guesses.

Note that the minimum around 20\% for the non-barrier height energetics is now very shallow, and that both functional
forms exhibit a deeper second minimum around 40\%. For the barrier heights, there only appears to be a single
minimum, at 42\% to be precise\cite{forty-two}. This value was retained for the final functionals.

The B97-1 like $\tau$-less form was frozen at this stage, and will henceforth
be denoted B97-K
(Becke-97 for `kinetics').

The $\tau$-dependent form was then subjected to its final 
refinement. On the one hand, we expanded the `training set' to an enlarged version of the HCTH/407 set\cite{HCTH3},
encompassing 464 energetic properties and 4008 Cartesian gradient components. 
This set includes nearly the full G3-99 set of molecules\cite{G3} minus some very large organic systems, plus
several dissociation energies of hydrogen-bonded complexes and stationary points on the ammonia dimer
potential surface taken from Ref.\cite{HCTH4}, and several ligand dissociation energies of transition metal
complexes --- in addition to the reaction barriers heights already discussed. On the other hand, we replaced TZ2P by the large basis set
described in the Computational Methods section.
Full details of all the data are available in the Supplementary Material\cite{supp}.

The final functional thus obtained is denoted BMK (from the present authors' initials and the K of `kinetics'). The parameters 
of both B97-K and BMK are given in Table \ref{tab2}.
Some of their values are somewhat surprising if we consider the coefficients of these
functionals when no barrier heights are included, as discussed above. We will first discuss the GGA (B97)
functional form. With a larger fraction of exact exchange, the zero-order correlation coefficients $c_{C\sigma\sigma,0}$ and
$c_{C\alpha\beta,0}$ will `try to correct back' the large amount of exchange mixing, together with a smaller local exchange
coefficient $c_{X\sigma,local,0}$. Introducing transition states into the fit to obtain B97-K will reverse this effect, since this
large amount of exact exchange is needed.

The $\tau$-HCTH form, however, behaves {\it the opposite way}. Here, with a larger
fraction of exact exchange, the non-local exchange coefficients can now `correct back' the exchange, yielding positive correlation
coefficients $c_{C\sigma\sigma,0}$ and $c_{C\alpha\beta,0}$.

When introducing the transition states into the set to obtain BMK, these
non-local coefficients are now needed for something else, and the zero-order correlation and exchange coefficients look similar to
the B97 functional with a larger fraction of exact exchange. While having a small effect on the other coefficients at the minimum in
Figure \ref{Fig1} (see Table II of reference \cite{tHCTH} for comparison), the kinetic energy density has a {\it very large} effect
as soon as the amount of exact exchange is not around 15-25\%. This explains the %observation which was reported before that
previously reported observation that
functionals employing the kinetic energy density are 
capable of simulating
exact exchange. It also implies that with large values
of the exact exchange mixing coefficient, the %improvements depending on functionals
benefits derived from incorporating
the kinetic energy density and similar
variables into DFT functionals will be even larger.

We shall now finally assess the new functionals for a variety of properties, in particular as to how they measure up compared
to other functionals designed either specifically for barrier heights or for equilibrium energetics.

\section{Assessment of Density Functionals}

As our final parametrization set of 464 systems is larger and arguably more diverse than any previously used, we shall use
it for comparison with other functionals as well. We have considered (a) the GGA functionals BP86 (Becke exchange\cite{B88X}
with Perdew 1986 correlation\cite{P86}), BLYP\cite{B88X,LYP}, HCTH/407\cite{HCTH3}, and PBE (Perdew-Burke-Ernzerhof\cite{PBE});
(b) the meta-GGA functionals $\tau$-HCTH\cite{tHCTH} and VSXC\cite{VSXC}; 
(c) the hybrid functionals B3LYP\cite{B3P91,LYP}, B97-1\cite{HCTH}, B97-2\cite{B97-2}, PBE0\cite{PBE,PBE0}, 
$\tau$-HCTH hybrid\cite{tHCTH}, KMLYP\cite{KMLYP} and mPW1K\cite{mPW1K}.
We employed the same large $spdf$ basis set for the assessment as was used for the final parametrization cycles of BMK.
Summary error statistics for all functionals, broken down by type of molecule or property, are presented in Table  \ref{tab3}.

For the non-hybrid functionals, the
results of the previous investigations with the TZ2P and cc-pVTZ basis sets\cite{basissetpaper} are corroborated: For all
molecules, the $\tau$-HCTH functional yields the lowest errors, followed by the HCTH/407 and VSXC functionals.
Particularly BP86 and PBE yield very large errors 
when considering geometries and atomization energies.
Interestingly, this pattern is not necessarily carried over from the TAEs to the reaction energies: for instance,
VSXC has half the RMS error for atomization energies of BLYP but both functionals perform similarly (somewhat poorly) for
reaction energies.
As to the geometry gradient errors 
which correlate with geometry errors, see also Refs.\cite{HCTH3,tHCTH}, the best performance appears to be
put in by 
those functionals that perform best for TAEs (i.e. the meta-GGAs and HCTH/407). 

HCTH/407 and both meta-GGA functionals in fact are overall competitive with the very popular B3LYP hybrid functional, in
turn
surpassed
by B97-1 and B97-2. Interestingly (although perhaps fortuitously), all hybrid functionals perform worse
than the (meta-)GGAs for transition metal ligation energies, although this problem is greatly mitigated in $\tau$-HCTH hybrid.
For our dataset, the best hybrid functionals (B97-1 and B97-2) overall represent an improvement over (meta-)GGAs for 
TAEs, but not significantly so for ionization potentials and electron affinities. As expected, the transition-state optimized
mPW1K and KMLYP functionals perform rather poorly for the thermochemical properties, although this would be mitigated somewhat
in the case of KMLYP by applying their recommended `high-level' additive correction. For geometric parameters, the RMS error
of mPW1K is twice, and that of KMLYP thrice, that of other functionals. Interestingly, mPW1K performs reasonably
well for the reaction energies. 

Our B97-K functional fares better than mPW1K, with errors for equilibrium energetics generally in between mPW1K and B97-1, 
and errors for geometries comparable to B3LYP or $\tau$-HCTH hybrid (cut in half compared to mPW1K). But it is for BMK that 
performance is particularly pleasing: its performance for most energetic properties is comparable to the best hybrid general-purpose
functionals, despite its high percentage of Hartree-Fock exchange. Indeed, for transition metal complexes it has the second
best performance of all hybrid functionals, after $\tau$-HCTH hybrid. For geometries, its RMS error is somewhat deteriorated
compared to B97-1 or to its closest kin, $\tau$-HCTH hybrid, but still much less so than that of mPW1K or KMLYP. The performance
of BMK for TAEs of neutral molecules is especially pleasing. Error statistics for ionization potentials and electron 
affinities are comparable to B3LYP. Overall, BMK can hold its own with general-purpose hybrid functionals
for equilibrium energetics and geometries.

Interestingly, the improvements previously reported\cite{tHCTH} for the $\tau$-HCTH hybrid functional compared to B97-1 now
vanish because of the larger basis set (compared to TZ2P used in the earlier work). This suggests again that some basis set
incompleteness error may have been absorbed in the parametrization of the HCTH family\cite{BMHcomment}.

For reaction barrier heights, the general-purpose hybrid functionals represent an improvement over all (meta-)GGAs except perhaps
$\tau$-HCTH. Yet they are clearly trumped by the transition state-optimized functionals, three of which (mPW1K, B97-K, and BMK)
have errors around 2 kcal/mol. 

Histograms of the errors for the BMK and B97-K functional, as well as for B3LYP and mPW1K, are
given in Figures S--1 through S--4 in the Electronic Supplement to this paper\cite{E-PAPS}. The error distributions
can be fitted rather well (correlation coefficients R ranging from 0.98 for B97-K to 0.993 for BMK)
by a normal distribution $A \exp(-(x-\mu)^2/2\sigma^2)$, which has been displayed in each graph
together with its mean $\mu$ and standard deviation $\sigma$. We can see that we did not merely
remedy an additive bias (with $\mu$ shifting from -4.1 kcal/mol for mPW1K to -0.6 kcal/mol for
BMK), but that the error distribution
was narrowed by a factor of two (from $\sigma$=6.7 kcal/mol for mPW1K to 3.3 kcal/mol for BMK), 
and that BMK indeed compares favorably to B3LYP 
($\mu$=-1.4, $\sigma$=4.3 kcal/mol) in both regards. In contrast, B97-K only represents a minor
improvement ($\mu$=-3.9, $\sigma$=6.4 kcal/mol) over mPW1K.

Overall, BMK turns in a very strong performance in Table \ref{tab3}, with energy errors in close proximity to hybrid functionals
that employ a much smaller fraction of exact exchange (15-25\%). The gradient and thus geometry errors are somewhat larger, and
probably in the range of a good meta-GGA or GGA functional such as HCTH/407.

Another important indication of the reliability of the functionals are frequency calculations. These are especially important in
order to find out if a functional has been `overfitted'. In this instance, the gradients would become quite small 
and achieve a small gradient error for the wrong reason:
The gradients at {\it any} geometry
would be small, even if far away from equilibrium. This would result in a flat potential energy surface and
thus in too low frequencies\cite{PhD}.
In order to ensure a balanced evaluation set (including organic molecules) and reliable data, we have mainly used  CCSD(T)/cc-pVQZ,
CCSD(T)/aug-cc-pVQZ, and CCSD(T)/ANO4321 harmonic frequencies from the literature (we may assume that these values
are quite accurate because of known error cancellation\cite{freq0} between core-correlation and post-CCSD(T) correlation 
effects)\cite{freq1,freq2,freq3,freq4,freq5,freq6,freq7,freq8,freq9,freq10,freq11,freq12,freq13}.
Using these reference values, we included the following molecules:
C$_2$H$_2$, C$_2$H$_4$, C$_6$H$_6$, CCl$_2$, CF$_2$, CF$_4$, CH$_2$NH, CH$_4$, CO, CO$_2$, F$_2$, FCCH, H$_2$, H$_2$CO, H$_2$O,
H$_2$SiO, HCN, HF, $cis$-HSiOH, $trans$-HSiOH, N$_2$, $cis$-N$_2$H$_2$, $iso$-N$_2$H$_2$, $trans$-N$_2$H$_2$, N$_2$O, NH$_3$,
SiH$_4$, SiF$_4$, SO$_2$, SO$_3$, and $s$-tetrazine. This makes for a total of 202 distinct harmonic frequencies: all computed values
are given in the Supplementary Material\cite{supp}, while mean signed,
mean absolute, and RMS errors over the whole set are displayed in Table \ref{tab4}. We investigated the performance of the BP86, BLYP, HCTH/407 and
PBE GGA functionals, together with B3LYP, B97-1, B97-2, PBE0, KMLYP, mPW1K, B97-K and BMK. For technical reasons, the BMK frequencies have been
calculated by numerical differentiation of gradients. 
For the harmonic frequencies investigated, GGA functionals generally underestimate, and hybrid
functionals with large fractions of exact exchange overestimate, the coupled-cluster values. Among the GGA functionals, HCTH/407
clearly yields the lowest errors, underestimating the frequencies by an average 30 cm$^{-1}$. There is little difference
between the hybrid functionals B3LYP, B97-1, B97-2 and PBE0, with an improvement of 17\% by B97-1 compared to B3LYP. Our
investigations\cite{freq13} indicate that this improvement for B97-1 %is largely based on 
%JM5 I guess you meant
largely derives from
organic molecules, which are described
even better than with B3LYP. Generally, the geometry gradient errors in Table \ref{tab3} seem to transfer rather well into
frequency errors for these functionals, with B97-K and BMK yielding errors comparable to a functional like HCTH/407. Again,
KMLYP and mPW1K yield frequency errors as bad or even worse than GGA functionals.

Hydrogen bonded complexes serve as another commonly used test for density functionals. 
Ab initio calculations on such systems converge quite rapidly with the electron correlation treatment (thanks to the long-range
character of the interaction studied) but very slowly with the basis set\cite{Halkier}, both with and without counterpoise\cite{cp} corrections.  
Since the use of extrapolation schemes almost eliminates the basis set superposition errors for atomization
energies\cite{W3} and hydrogen-bonded complexes such as the ammonia dimer\cite{HCTH4}, it is safe to assume that use of standard
W2 theory\cite{W2} will suffice. Further touchstones for hydrogen-bonded complexes include the lengths of the
hydrogen-bonds and hydrogen-bonded shifts calculated using CCSD(T)/aug'-cc-pVTZ. The harmonic frequency shifts are obtained at the
MP2/aug'-cc-pVTZ level.
We have computed reference values for nine hydrogen bonded complexes which were also used in a previous
study\cite{TBH}, in addition to the ammonia dimer. All these values (Table \ref{tab5}), with the exception of the MP2 frequency shifts (for which we
would expect a deviation of about 15\%), should be very accurate. In Table \ref{tab6}, the mean \% and RMS \% errors are defined
by: ${\rm mean} = \rm \frac{1}{10} \sum_i^n \left(\frac{value(complex(i))}{value(reference(i))}-1\right) \times 100\,\%$
and ${\rm RMS} = \rm \sqrt{\frac{1}{10}\sum_i^n \left(\frac{value(complex(i))}{value(reference(i))}-1\right)^2}
 \times 100\,\%$. In this Table, all interaction energies (MP2/a'pVQZ and DFT/aug-TZ) have been counterpoise-corrected. For the DFT
and MP2 methods tested, counterpoise corrections are generally on the order of about 0.5-2.0 kJ/mol, which is sometimes still quite large
relative to the interaction energies being considered.
For the reference W2 values, {\it all} residual basis set superposition error
(BSSE) is likely to have been
eliminated. Thus, we would expect the mean error to change by +2-3\%, so that the mean percentual error of
MP2 in Table \ref{tab5} will be approximately zero, and the functionals {\it underestimating} the interaction energy (like HCTH/407
and B3LYP) will become more accurate. All the computed values can be found in the supplementary
material\cite{supp}. Generally, it is believed that MP2 renders much more accurate results than density functional theory for
hydrogen-bonded systems.
Overall, the {\it consistency} of results of MP2 is only matched by one functional, surprisingly B97-K. The errors of this
functional
are low for all tested properties, from the dissociation energy to the frequency shift
of the hydrogen bond. BMK, on the other hand, renders a mediocre performance, which looks quite similar to the B97-2 functional.
Concerning the dissociation energies, HCTH/407 is the only functional giving an RMS \% error of less than 10.
Only the hydrogen bond distance itself is consistently overestimated by about 4\%. The same is true for B97-1: For all properties
investigated it yields low errors, except for the geometry shift. Generally, hybrid functionals are
more accurate than GGA functionals (with HCTH/407 yielding particularly good dissociation energies and BLYP hydrogen bond lengths),
which are in turn much more accurate than the other two `kinetics' functionals investigated. mPW1K shows the behavior we
would expect from the HF method, i.e., underestimating the interaction energy, the hydrogen and frequency shifts by about 40\%,
and overestimating the bond length. KMLYP on the other hand shows some behavior normally associated with
a functional fit to
the uniform electron gas, such as LSDA: The interaction energy and the shifts are overestimated and the bond length underestimated.
Overall, the new functionals show much improved performance for hydrogen bonded complexes, with 
B97-K, surprisingly, clearly yields the lowest error of all functionals tested.

Finally, we will consider performance of BMK and other functionals for transition states outside our training set. 
Unlike for the latter, all transition states will be optimized with the functional and basis set used. The results are 
presented in detail in Table S-1 in the Electronic Supplement\cite{E-PAPS}: summary error
statistics can be found in Table \ref{tab8}.
The source of the reference values was indicated in the second column of Table S-1. Entries 1--24 are the reactions from
the training set. The remainder are a large subset of the reactions considered by Kang and Musgrave in the KMLYP paper\cite{KMLYP},
and finally the reactions from the recent paper of Houk and coworkers\cite{Houk} on pericyclic reactions
like ring closings, hydrogen shifts and Diels-Alder reactions. For this class of reactions, Houk and coworkers found average errors
ranging from 1.7 to 3.2 kcal/mol
for several hybrid functionals. It should be noted that the average uncertainty of the experimental
values is itself 0.7 kcal/mol, casting some doubt on the validity of overly detailed comparisons for these systems. In
one case (the ring closing of 1,3-butadiene to cyclobutene), the 0.2 kcal/mol stated uncertainty appears to be somewhat 
optimistic: our computed W1 barrier deviates by 1.8 kcal/mol
changing by at most 0.1 kcal/mol when computed
at different calculated reference geometries, despite the absence of significant nondynamical correlation. The other
reactions we investigated were the hydrogen shifts of 1,3-cyclopentadiene and 1,3 pentadiene, again with the result that
the experimental values might have larger deviations than their given uncertainties: For the former we found a difference of
2.2 kcal/mol (which is close to their CBS-GB3 number reported), compared to the stated uncertainty of 0.5 kcal/mol.
For the latter, the W1 result is somewhat closer to experiment with a deviation of 0.7 kcal/mol, the reported uncertainty being
0.5 kcal/mol. In all three cases, the experimental barriers are lower than our computed values.

In addition, the experimental numbers of Kang and Musgrave (marked with footnote symbol $^a$) deviate
significantly from the ones reported by Truhlar and co-workers for reactions 19--24. This casts some doubt on the accuracy of the former:
However, since we do not have any other data available for many reactions, we will still use this set for comparison.

In Table S-1, we have tested mPW1K, KMLYP, B3LYP, B97-1, and B97-2 together with the two newly obtained density
functionals and the $spdf$ basis set combination described in the Computational Methods section. 
We will refrain from the use of small basis sets in order to calculate transition states, 
in order not to blur our comparison with 
additional error sources such as BSSE and basis set incompleteness effects on reference geometries.
During pre-optimizations, we in fact found that in some
difficult cases, we were unable to locate a transition state with a basis set like 6-31+G*.
Overall, the functionals developed for
transition states show a strong performance throughout for the barriers. The B97-1 and B3LYP functionals yield very inconsistent results: While some reaction barriers are actually overestimated, 
many transition states cannot be found at all, lie below the reactants in energy (albeit above the entry 
channel long distance complex: examples were previously reported for S$_N$2 reactions\cite{mPW91,ParthiSN2}),
or are underestimated by up to 10 kcal/mol (e.g. for the transition state of
the CH$_3$F + H $\rightarrow$ CH$_3$ + HF reaction, \# 63).
For most reactions, B97-1 does even worse than B3LYP, usually underestimating barriers by an additional kcal/mol. It is somewhat
peculiar that the forward reactions of the type AH + H $\rightarrow$ A + H$_2$ show a lower error for B97-1 (reactions 9--15,
36, 38, 39, 53, 63--65, and 70) than for B3LYP, with 
the opposite effect seen for the reverse reactions of this type (reactions 6, 7, 19, 20, and 49).
B97-2 usually predicts somewhat higher barriers than B97-1.

We will now turn to the summary error statistics in Table \ref{tab8}.
Compared to Table \ref{tab3} --- where we calculated single-point energies at the QCISD geometries --- the RMS
errors of all functionals are minimized further. The BMK functional yields the lowest mean absolute error of 1.1 kcal/mol, showing
a similar performance to B97-K and mPW1K. KMLYP yields results between the functionals which have 20\% (B3LYP, B97-1) exact exchange
and the ones having around 42\% (BMK, B97-K, and mPW1K).
While B97-2 represents an improvement over B97-1 (as noted in Ref.\cite{PCCP2004}), performance for barrier heights
is still clearly inferior to BMK, as well as to B97-K and mPW1K. (B97-2 and KMLYP are somewhat
more closely matched.)

For the fourth set of Houk and co-workers in Table \ref{tab8}, we reported error statistics to either ten (25--34) or eight (25--32)
reactions because of the 
discrepancies of the various computed values among each other and with experiment for reactions 33--35, particularly the last one.
For this set, B97-1 is the only functional that underestimates the transition barriers on average and shows the lowest RMS error.
The RMS errors of BMK and mPW1K are worse by about 0.7 kcal/mol, followed by B3LYP, B97-K and KMLYP.

Finally, the results obtained for our training set (reactions 1--24 in Table S-1) are further validated by comparing
to Kang and Musgrave's experimental numbers and
their CBS-APNO calculations. Here, KMLYP yields errors comparable to the other functionals developed for transition states,
probably because they used this set for determining their functional. Again, B3LYP and B97-1 show poor performance, B97-2 less so. As Kang
and Musgrave have already reported, G2 also yields very large errors, overestimating the experimental numbers
by an average of 3 kcal/mol. For transition states, G2 seems to be less accurate than the functionals tested.
This should not surprise the reader, considering the heavy reliance on low-order perturbation
theory --- which often has problems with transition states because of its hypersensitivity to nondynamical
correlation effects --- in the G2 method.

Summarizing this section, without the two new functionals one is in a dilemma about whether to 
compromise on the accuracy of equilibrium energetics (with transition-state biased functionals
like mPW1K and KMLYP, compare Tables
\ref{tab3}, \ref{tab4} and \ref{tab6} with Tables \ref{tab8} and S-1)
or on that of barrier heights (with traditional
general-purpose functionals). B97-K and in particular BMK allow users to `have their cake and 
eat it too'. In fact, overall performance
of BMK for all properties considered here is such
that it can legitimately be considered a general-purpose functional, albeit with its applicability
extended to transition states.

\section{Conclusions}

Prior to this work, all exchange-correlation functionals that consistently perform reliably for transition states and barrier heights
(such as BH\&HLYP, mPW1K and KMLYP)
achieve this feat at the expense of seriously degraded performance for all other properties.
We have developed two new exchange-correlation functionals suitable for transition states, the hybrid GGA functional B97-K 
(Becke-97 for kinetics) and the hybrid meta-GGA functional BMK (Boese-Martin for kinetics). These functionals, especially BMK,
were parametrized against very large and diverse training sets. Both functionals achieve performance
similar to mPW1K for transition states. However, B97-K and especially BMK perform much more reliably for more general properties
than
those earlier specialized functionals. Overall performance of B97-K is intermediate between mPW1K and the very reliable
B97-1 functional (except for perhaps fortuitously excellent performance for hydrogen bonds);
the BMK functional actually outperforms B3LYP for most properties and is competitive with the best currently
available functionals for equilibrium energetics. Its performance for geometries and vibrational frequencies is somewhat less
good than B3LYP but markedly superior to all available specialized functionals for transition states. Unlike these latter,
BMK can actually be considered a reliable general-purpose functional whose capabilities have been expanded to cover 
transition states.

To computational chemists engaged in mechanistic studies, BMK offers a single functional capable of describing all 
aspects of the potential surface (energetics, kinetics, structures,...) Initial experiments for electrical properties
(to be reported on in a future paper) yield promising results as well, in part because self-interaction error issues 
are less acute for functionals with high percentages of Hartree-Fock exchange.

The surprisingly good performance of BMK appears to hinge on the combination of a high percentage of Hartree-Fock 
exchange with terms dependent on the kinetic energy density. The latter appears to afford a `back-correction' for
excessive Hartree-Fock exchange in systems where it is undesirable, and in effect to simulate variable exact 
exchange to some extent.
Some evidence exists that the same properties of BMK make it more suitable for describing transition metal complexes 
than other hybrid functionals. These observations should shed some light on the problem of developing both more accurate and 
more widely applicable functionals than currently available.

In sum, we believe we have added a very useful tool to the computational chemist's toolbox. 

\section*{Note added in revision}

A referee drew our attention to a recent paper by Zhao et al.\cite{NZJAH}
in which a successor to mPW1K, denoted BB1K (Becke-Becke 1-parameter for kinetics) was proposed. This functional combines the Becke 1988 GGA exchange functional\cite{B88X} with Hartree-Fock exchange and Becke's 1995 meta-GGA correlation functional\cite{B95}. The percentage of Hartree-Fock exchange is
empirically adjusted to minimize the RMS error over a small dataset of nine barrier
heights, the optimum value turning out to be 42\% like in the present BMK functional. Clearly, this functional was not as broadly and comprehensively
parametrized as BMK or even B97-K. Furthermore, Becke himself\cite{B97} states:
"The functional of Part IV [i.e., Ref.\cite{B95}] is problematic in very weakly
bound systems... and is therefore not recommended. This will be corrected in
future publications." In summary, we have reason to believe that the present BMK functional will be considerably more reliable and widely applicable than the one proposed in Ref.\cite{NZJAH}.

\begin{acknowledgments}
ADB acknowledges a postdoctoral fellowship from
the Feinberg Graduate School (Weizmann Institute), and is grateful to W. Klopper for being able to finish this
work at the Forschungszentrum Karlsruhe. Research at Weizmann
was supported by the Minerva Foundation, Munich, Germany, by the Lise Meitner-Minerva Center for
Computational Quantum Chemistry (of which JMLM is a member), and by the
Helen and Martin Kimmel Center for Molecular Design. 
The authors also thank Mark A. Iron for helpful discussions.
\end{acknowledgments}

\newpage

\newpage

\newpage

\newpage

\clearpage
\noindent
\section*{Figure Captions}
{\bf Figure 1}:
Dependence on the exact exchange mixing coefficient of the RMS error
for the B97 and $\tau$-HCTH forms (energetic properties of the 147 set, kcal/mol).\\

{\bf Figure 2}:
Dependence on the exact exchange mixing coefficient, and the weight assigned to reaction barrier heights in the parametrization, of the RMS error for the B97 form (energetic properties of the 147 set, kcal/mol).\\

{\bf Figure 3}:
RMS error (kcal/mol) for barrier heights with the B97 form as a function of the exact exchange admixture and the
weight assigned to barrier heights in the parametrization.\\

{\bf Figure 4}:
RMS errors (kcal/mol) of the B97 and $\tau$-HCTH forms for the extended 147 set and for barrier heights as a function of the exact exchange admixture and the
weight assigned to barrier heights in the parametrization.\\

\newpage
\clearpage
\begin{table}
\caption{W1 and W2 reaction barrier heights (kcal/mol) computed in this study, compared to experimental data\cite{mPW1Keval}\label{tab1}}
\begin{tabular}{|l|l l l|l l l|} \hline\hline
Reaction                    &\multicolumn{3}{c|}{forward barrier}&\multicolumn{3}{c|}{reverse barrier}\\ \hline
                                                       &  W1   &  W2   & exp. &  W1   & W2    & exp.  \\ \hline
 C$_5$H$_8$          $\rightarrow$ C$_5$H$_8$          & 36.80 &       & 38.4 & 36.80 &       &  38.4 \\ 
 C$_2$H$_6$ + NH     $\rightarrow$ C$_2$H$_5$ + NH$_2$ & 19.35 &       & 18.4 &  9.86 &       &   8.0 \\ 
 C$_2$H$_6$ + NH$_2$ $\rightarrow$ C$_2$H$_5$ + NH$_3$ & 11.25 &       & 10.4 & 17.93 &       &  17.8 \\ 
 C$_2$H$_6$ + OH     $\rightarrow$ C$_2$H$_5$ + H$_2$O &  3.52 &       &  3.4 & 20.49 &       &  20.7 \\ 
 CH$_3$     + H$_2$  $\rightarrow$ CH$_4$     + H      & 11.89 & 11.89 & 12.1 & 14.88 & 14.92 &  15.0 \\ 
 CH$_3$     + OH     $\rightarrow$ CH$_4$     + O      &  9.00 &       &  7.8 & 14.28 &       &  13.7 \\ 
 CH$_4$     + NH     $\rightarrow$ CH$_3$     + NH$_2$ & 21.98 &       & 22.7 &  8.95 &       &   8.4 \\ 
 CH$_4$     + NH$_2$ $\rightarrow$ CH$_3$     + NH$_3$ & 13.92 &       & 15.5 & 16.97 &       &  17.9 \\ 
 CH$_4$     + OH     $\rightarrow$ CH$_3$     + H$_2$O &  6.22 &       &  6.7 & 19.66 &       &  20.2 \\ 
 H$_2$      + Cl     $\rightarrow$ H          + HCl    &  6.97 &  7.90 &  8.7 &  5.62 &  4.98 &   5.6 \\ 
 H$_2$      + F      $\rightarrow$ H          + HF     &  (a) &       &  1.8 &       &       &  33.2 \\ 
 H$_2$      + OH     $\rightarrow$ H          + H$_2$O &  5.29 &  5.40 &  5.7 & 21.72 & 21.65 &  22.0 \\ 
 H$_2$S     + H      $\rightarrow$ SH         + H$_2$  &  3.45 &  3.62 &  3.6 & 17.01 & 17.14 &  17.4 \\ 
 H$_2$      + H      $\rightarrow$ H          + H$_2$  &  9.60 &  9.66 &  9.6 &  9.60 &  9.66 &   9.6 \\ 
 CH$_3$OH   + H      $\rightarrow$ CH$_2$OH   + H$_2$  &  9.64 &       &  7.3 & 15.70 &       &  13.8 \\ 
 CH$_3$     + HCl    $\rightarrow$ CH$_4$     + Cl     &  1.37 &       &  1.8 &  7.39 &       &   7.8 \\ 
 HCl        + H      $\rightarrow$ H          + HCl    & 17.21 & 17.25 & 18.0 & 17.21 & 17.25 &  18.0 \\ 
 N$_2$H$_2$ + H      $\rightarrow$ N$_2$H     + H$_2$  &  2.90 &  2.81 &  5.9 & 40.48 & 40.64 &  41.0 \\ 
 PH$_3$     + H      $\rightarrow$ PH$_2$     + H$_2$  &  2.58 &  2.78 &  3.2 & 24.58 & 24.75 &  25.5 \\ 
 NH$_3$     + OH     $\rightarrow$ NH$_2$     + H$_2$O &  3.54 &       &  3.2 & 13.92 &       &  13.2 \\ 
 OH         + Cl     $\rightarrow$ O          + HCl    &  (b) &       &  9.8 &       &       &   9.9 \\ 
 OH         + H      $\rightarrow$ O          + H$_2$  & 10.96 & 10.94 & 10.1 & 13.25 & 13.48 &  13.1 \\ \hline\hline
\end{tabular}

\vspace{12pt} 

(a) CCSD(T) of doubtful value due to severe nondynamical correlation. Value of 2.31 kcal/mol taken from benchmark calculation
by K\'allay et al.\cite{KallayHF} \\
(b) CCSD(T) of doubtful value due to severe nondynamical correlation. Experimental value used (see text).\\

\end{table}

\newpage
\clearpage

\begin{table}
\caption{Coefficients of the new functionals, where both new functionals have 42\% exact exchange. In addition, the B97 and 
$\tau$-HCTH functionals when just fitting them to the 147 set (without transition states) are shown for comparison.\label{tab2}}
\begin{tabular}{@{}|l|c c|c c|@{}} \hline \renewcommand{\arraystretch}{2}
Functional for                   & \multicolumn{2}{c|}{transition states + energies} & \multicolumn{2}{c|}{energies} \\ \hline\hline
Name                             & B97-K     & BMK        & B97-K like        & BMK-type \\ \hline
Exact exchange mixing            & \multicolumn{2}{c|}{42\%}&  \multicolumn{2}{c|}{45\%}\\ \hline
$c_1=c_{X\sigma,local,0}$        &  0.507863 &   0.474302 &   0.454579 &   0.554540  \\ \hline
$c_2=c_{X\sigma,non-local,0}$    &           &  -0.192212 &            &  -0.808213  \\ \hline
$c_3=c_{C\sigma\sigma,0}$        &  0.12355  &  -2.19098  &  -2.51595  &   0.45226   \\ \hline
$c_4=c_{C\alpha\beta,0}$         &  1.58613  &   1.22334  &   1.68137  &   0.89243   \\ \hline
$c_5=c_{X\sigma,local,1}$        &  1.46873  &   2.77701  &   1.70057  &   0.11565   \\ \hline
$c_6=c_{X\sigma,non-local,1}$    &           &   0.47394  &            &   0.68619   \\ \hline
$c_7=c_{C\sigma\sigma,1}$        &  2.65399  &  23.8939   &  12.8997   &  -9.17579   \\ \hline
$c_8=c_{C\alpha\beta,1}$         & -6.20977  &  -3.4631   &  -8.9966   &   6.47624   \\ \hline
$c_9=c_{X\sigma,local,2}$        & -1.51301  & -11.4230   &  -2.2741   &  -0.32078   \\ \hline
$c_{10}=c_{X\sigma,non-local,2}$ &           & -26.6188   &            & -16.0399    \\ \hline
$c_{11}=c_{C\sigma\sigma,2}$     & -3.20694  & -44.3303   & -11.0840   &  27.6665    \\ \hline
$c_{12}=c_{C\alpha\beta,2}$      &  6.46106  &  10.0731   &  12.6502   & -36.9632    \\ \hline
$c_{13}=c_{X\sigma,local,3}$     &           &  11.7167   &            &  -0.6402    \\ \hline
$c_{14}=c_{X\sigma,non-local,3}$ &           &  22.4891   &            &  -5.7755    \\ \hline
$c_{15}=c_{X\sigma\sigma,3}$     &           &  22.5982   &            & -19.9035    \\ \hline
$c_{16}=c_{X\alpha\beta,3}$      &           & -11.1974   &            &  31.8788    \\ \hline\hline
\end{tabular}
\end{table}

\newpage
\clearpage

\begingroup
\squeezetable
\begin{table}
\caption{RMS energy errors for the new functionals, compared with  BP86, BLYP, HCTH/407, PBE, $\tau$-HCTH, VSXC, B3LYP,
B97-1, B97-2, PBE0, $\tau$-HCTH hybrid, KMLYP and mPW1K.\label{tab3}}
\begin{tabular}{|l|l|l l l l l l l l l l l l l l l|} \hline\hline
\multicolumn{2}{|l|}{Class                      }  & BP86 & BLYP & HCTH & PBE & $\tau$- & VSXC &
B3LYP & B97-1 & B97-2 & PBE0 & $\tau$-       & KMLYP & mPW1K & B97-K & BMK  \\  
\multicolumn{2}{|l|}{                     }  &   &  &/407  &   & HCTH &   &
  &  &   &   & HCTHh       &   &  &   &    \\ \hline
\multicolumn{2}{|l|}{All systems     (404)      }  & 21.2 & 10.3 &     7.9  & 18.9 & 7.1        & 7.8  &7.7   & 5.3   & 5.3   & 9.3  &      5.5           & 25.6$^*$& 14.0 & 10.3 & 5.6  \\ 
\multicolumn{2}{|l|}{Neutrals (219)      }  &  21.8 & 11.6 &     8.6  & 21.3 & 7.6        & 6.8  &8.1   & 4.8   & 4.6   & 8.5  &      5.1           & 27.7$^*$& 15.8 & 11.5 & 4.5  \\ 
\multicolumn{2}{|l|}{Anions  (58)      }   & 15.2 &  9.5 &     7.8  & 13.4 & 6.4        & 10.3 &9.1   & 6.3   & 6.1   & 9.7  &      6.3           & 20.9$^*$& 13.9 & 10.8 & 7.9  \\ 
\multicolumn{2}{|l|}{Cations (88)      }   & 15.1 &  8.3 & 7.0      & 18.8 & 6.3        &  9.3 &5.7   & 5.5   & 5.6   & 11.8 &      5.6           & 27.7$^*$& 10.2 &  7.0 & 6.6  \\ 
\multicolumn{2}{|l|}{TM complexes (5)      }  & 11.1 &  4.2 & 4.7      & 13.3 & 4.1        &  5.5 &17.0  & 13.1  & 15.2  & 14.3 &      7.2           & 29.0$^*$& 28.7 & 25.0 & 10.8 \\ 
\multicolumn{2}{|l|}{TSes (24)     }  &  10.1 &  8.5 &  7.0     & 10.3 & 7.9        &  6.2 & 5.0  & 5.2   & 4.1   & 4.9  &      6.1           & 3.2     & 1.8  & 2.0  & 2.0 \\ 
\multicolumn{2}{|l|}{React. ener. (62)     }  &  5.0 &  6.4 &  5.7     &  5.3 & 4.6        &  6.8 & 4.4  & 4.2   & 3.4   & 4.3  &      4.0           & 8.2$^*$ & 4.8  & 4.4  & 3.7 \\ \hline
\multicolumn{2}{|l|}{IPs (80) }  &  0.23 & 0.28 & 0.25     & 0.24 & 0.23       & 0.24 &0.23  & 0.22  & 0.21  & 0.23 &     0.22           & 0.41$^*$& 0.27 & 0.27 & 0.24 \\ 
\multicolumn{2}{|l|}{EAs   (58) }   & 0.22 & 0.18 & 0.24     & 0.19 & 0.18       & 0.18 &0.19  & 0.16  & 0.18  & 0.20 &     0.17           & 0.27$^*$& 0.26 & 0.22 & 0.19 \\ \hline
\multicolumn{2}{|l|}{$\sum\Delta$gradient (4008)}&  17.52 & 19.01 & 12.10  & 16.84 & 11.56     & 10.96 &11.17 & 10.72 & 11.80 & 12.33 &   10.77          & 30.71     & 20.24 & 11.61 & 13.53 \\ \hline
\end{tabular}
\end{table}
\vspace{12pt}\noindent All errors in kcal/mol, except ionization potentials (eV), electron affinities (eV), and sum of gradient components (a.u.) at
their reference (experimental/theoretical)
equilibrium geometry.\\
(*) Exclusive of `high level correction' from Ref.\cite{KMLYP}.\\
\endgroup

\newpage
\clearpage

\begin{table}
\caption{Harmonic frequency errors (cm$^{-1}$) for a variety of functionals.\label{tab4}}
\begin{tabular}{|l|r c r|} \hline\hline
Functional & mean error & mean absolute error & RMS error \\ \hline
BP86       & -57.6      & 58.9                & 79.1      \\
BLYP       & -61.5      & 62.2                & 83.9      \\
HCTH/407   & -27.8      & 35.4                & 52.0      \\
PBE        & -54.0      & 55.9                & 75.0      \\
B3LYP      &  -2.7      & 21.6                & 30.8      \\
B97-1      &  -2.5      & 18.3                & 26.3      \\
B97-2      &  12.6      & 22.9                & 32.8      \\
PBE0       &  12.5      & 23.9                & 34.8      \\
KMLYP      &  86.3      & 87.0                & 104.6     \\
mPW1K      &  57.6      & 58.0                & 72.6      \\
B97-K      &  28.4      & 30.2                & 40.7      \\
BMK        &  27.3      & 31.7                & 45.9      \\\hline\hline
\end{tabular}
\end{table}

\newpage
\clearpage
\begingroup
\squeezetable
\begin{table}
\caption{Reference values for the ten hydrogen-bonded complexes.\label{tab5}}
\begin{tabular}{|l|l|l|l|l|} \hline\hline
Dimer        &  $D_e$& Donor $\Delta r_{XH}$&  H-Bond Length&  Freq. Shift \\
             & (kJ/mol)   &  (\AA)        &   (\AA)       & (cm$^{-1}$) \\\hline
(ClH)(NH$_3$)    & 34.9   & 0.0434        &  1.79         &  -770 \\
(CO)(HF)         &   7.1  & 0.0018        &  2.07         &   -32 \\
(FH)(NH$_3$)     &  52.1  & 0.0317        &  1.70         &  -785 \\
(H$_2$O)$_2$     &  20.8  & 0.0063        &  1.95         &  -170 \\
(H$_2$O)(NH$_3$) &  26.8  & 0.0117        &  1.98         &  -298 \\
(H$_3$O)(H$_2$O)$^+$ (*)& 141.2&0.2366        &  1.19         & -2780 \\
(HCl)$_2$        &   8.4  & 0.0039        &  2.56         &   -72 \\
(HF)$_2$         &  19.1  & 0.0057        &  1.82         &  -138 \\
(OC)(HF)         &  14.8  & 0.0061        &  2.08         &  -172 \\
(NH$_3$)$_2$     &  13.2  & 0.0033        &  2.30         &  -134 \\\hline\hline
\end{tabular}
\vspace{12pt}

\noindent Dissociation energies by W2 theory; geometry shifts CCSD(T)/aug$'$-cc-pVQZ; frequency shifts MP2/aug$'$-cc-pVQZ.
aug$'$-cc-pVQZ stands for cc-pVQZ on hydrogen and aug-cc-pVQZ on all other elements.\\

\noindent (*) Symmetric structure with central H atom, better written as H$_5$O$_2^+$.
\end{table}
\endgroup

\newpage
\clearpage

\begin{table}
\caption{Relative errors of the functionals and MP2/aug$'$-cc-pVQZ for the hydrogen-bonded complexes.\label{tab6}}
\begin{tabular}{|l|r r|r r|r r|r r|} \hline\hline
           &\multicolumn{2}{c|}{Dissoc. energy}&\multicolumn{2}{c|}{$\Delta r$(H-bond)}&\multicolumn{2}{c|}{$\Delta r$(H-bond)}
&\multicolumn{2}{c|}{Freq. Shift} \\\hline
Functional & mean \% & RMS \% & mean \% & RMS \% &  mean \% & RMS \% &  mean \% & RMS \% \\ \hline
BP86       & -12.5   & 25.3   &  80.9   &  91.2  &   -2.87  &  4.20  &  40.3    & 49.9   \\
BLYP       & -14.6   & 21.6   &  42.7   &  46.8  &   -1.01  &  1.97  &  16.9    & 30.2   \\
HCTH/407   &  -7.5   &  9.0   &  -2.7   &  26.3  &    4.28  &  5.88  & -14.1    & 23.3   \\
PBE        &   8.6   & 14.0   &  76.9   &  86.0  &   -3.30  &  4.54  &  40.1    & 49.3   \\
B3LYP      &  -8.5   & 15.0   &  29.0   &  31.3  &   -0.39  &  1.32  &   9.1    & 16.3   \\
B97-1      &   3.0   &  6.0   &  33.4   &  38.2  &   -1.01  &  2.07  &  11.4    & 16.2   \\
B97-2      & -19.6   & 25.3   &  18.9   &  25.0  &    1.08  &  2.96  &   4.4    &  8.5   \\
PBE0       &   2.3   &  9.4   &  46.1   &  52.8  &   -2.46  &  3.26  &  24.0    & 29.3   \\
KMLYP      &  32.2   & 34.5   &  51.4   &  57.6  &   -5.38  &  5.66  &  33.0    & 46.0   \\
mPW1K      & -38.0   & 40.9   & -32.1   &  36.3  &    9.67  & 10.55  & -32.9    & 40.0   \\
B97-K      &  -1.7   &  6.7   &   2.8   &   6.0  &    1.26  &  1.40  &  -9.2    & 13.0   \\
BMK        & -17.8   & 24.8   &  32.4   &  37.9  &   -0.68  &  1.50  &   4.5    & 25.2   \\\hline
MP2        &  -1.8   &  8.0   &  15.2   &  19.6  &   -0.33  &  1.47  &  (0.0)   & (0.0)  \\\hline\hline
\end{tabular}
\end{table}

\newpage
\clearpage

\begingroup
\squeezetable
\begin{table}
\caption{Error statistics of the functionals for the various sets of barrier heights in Table
S-1.\label{tab8}}
\begin{tabular}{|l|l l l|l l l l l l l | l|} \hline\hline
&\multicolumn{3}{c|}{Set Specifications}& \multicolumn{8}{c|}{Method}
\\ \hline
Set & Reference & \# TSs & Error     & BMK    & B97-K  & mPW1K & KMLYP &
B3LYP & B97-1 & B97-2 & G2 \\ \hline
1,2 & W1/W2,exp &  24    & mean      & -0.8   & -0.3   & -0.7  &  -2.4 &
-4.1 & -3.9 & -2.6 & \\
1,2 & W1/W2,exp &  24    & mean abs. &  1.1   &  1.3   &  1.3  &   2.5 &
 4.1 &  4.1 & 3.0 & \\
1,2 & W1/W2,exp &  24    & RMS       &  1.5   &  1.6   &  1.5  &   2.9 &
 4.3 &  4.6 & 3.5 & \\ 
1,2 & W1/W2,exp &  24    & max(+)    &  1.5   &  2.0   &  1.7  &   0.9 &
-1.7 &  0.8 & 1.8 & \\ 
1,2 & W1/W2,exp &  24    & max(--)    & -3.7   & -3.0   & -3.2  &  -5.0 &
-6.9 & -7.0 & -6.4 & \\ \hline
  4 & W1,exp    &  10(8) & mean      &1.7(1.9)&3.2(2.9)&2.2(1.1)&3.4(1.9)&
1.3(0.8)&-1.0(-1.1) & 0.9(-0.1)& \\
  4 & W1,exp    &  10(8) & mean abs. &2.9(2.2)&3.4(2.9)&2.9(2.0)&3.9(2.5)&
3.3(2.2)&2.2(1.4) & 2.4(1.9)& \\
  4 & W1,exp    &  10(8) & RMS       &3.5(2.7)&4.4(3.5)&3.6(2.3)&5.1(3.2)&
4.4(2.8)&2.9(1.7) & 3.3(2.3)& \\
  4 & W1,exp    &  10(8) & max(+)    &6.8(5.1)&9.9(6.1)&6.9(4.5)&10.2(6.1)&
10.6(5.3)&5.1(0.9) & 8.4(3.5)& \\
  4 & W1,exp    &  10(8) & max(--)&-4.3(-0.7)&-0.9(0.4)&-2.0(-2.0)&-1.2(-1.2)&
-4.5(-2.2)&-6.0(-3.0) & -3.1(-3.1)& \\\hline
  3 & CBS-APNO  &  22    & mean      &  0.0   &  0.4   &  0.5  &  -0.8 &
-4.1 & -4.0 & -2.6 & 2.1 \\
  3 & CBS-APNO  &  22    & mean abs. &  1.3   &  1.7   &  1.5  &   1.8 &
 4.3 &  4.2 & 2.9 & 2.2 \\
  3 & CBS-APNO  &  22    & RMS       &  1.7   &  2.2   &  2.0  &   2.3 &
 4.9 &  4.9 & 3.5 & 2.8 \\
  3 & CBS-APNO  &  22    & max(+)    &  5.0   &  5.4   &  5.5  &   4.8 &
 2.1 &  1.5 & 1.6 & 8.7 \\
  3 & CBS-APNO  &  22    & max(--)    & -2.8   & -2.8   & -2.9  &  -4.6 &
-9.2 & -8.5 & -6.7 &-1.1 \\\hline
  3 & exp       & 32(31) & mean      &  0.7   &  0.9   &  1.3  &  -0.8 &
-3.3 & -3.3 & -1.5 & (2.9) \\
  3 & exp       & 32(31) & mean abs. &  1.7   &  1.7   &  1.4  &   1.8 &
 3.7 &  3.8 & 3.0 & (3.3) \\
  3 & exp       & 32(31) & RMS       &  2.2   &  2.4   &  1.9  &   2.3 &
 4.6 &  4.6 & 3.5 & (4.2) \\
  3 & exp       & 32(31) & max(+)    &  5.6   &  5.9   &  5.7  &   3.1 &
 2.2 &  2.6 & 3.8 & (13.7)\\
  3 & exp       & 32(31) & max(--)    & -2.4   & -2.5   & -0.8  &  -2.5 &
-11.6& -7.9 & -5.8 & (-3.3)\\\hline\hline
\end{tabular}
\end{table}
\endgroup

\noindent Set 1 consists of the reactions from Table \ref{tab1}; Set 2 are some anionic S$_N$2 reactions taken from Ref.\cite{ParthiSN2}; Set 3 was taken from Kang and Musgrave\cite{KMLYP}; the pericyclic
organic reactions from the work of Houk and coworkers\cite{Houk} constitute Set 4.

%\end{document}
%%JM HERE ENDS PAPER PROPER
\clearpage

\setcounter{page}{0}
\renewcommand{\thepage}{S--\arabic{page}}
{\LARGE E-PAPS Supplementary data for paper}
\vspace{3cm}\\
\thispagestyle{empty}
%%JM HERE BEGINS E-PAPS STUFF MINUS TITLE PAGE

\noindent Will eventually be available online at (conjectured URL)\\
\url{http://netserver.aip.org/cgi-bin/epaps?ID=JCPSA6-121-303431}\\
A mirror copy will be kept at 
\url{http://theochem.weizmann.ac.il/web/papers/BMK.html}\\

\vspace{1in}

\noindent Note that in the present preprint, Table S-1 has been renumbered Table VIII to avoid 
conflicts with the \url{arXiv.org} PDF hyperlinking technology. 
                                                                          
\clearpage

\clearpage
\begin{figure*}
\includegraphics[height=12cm,width=18cm]{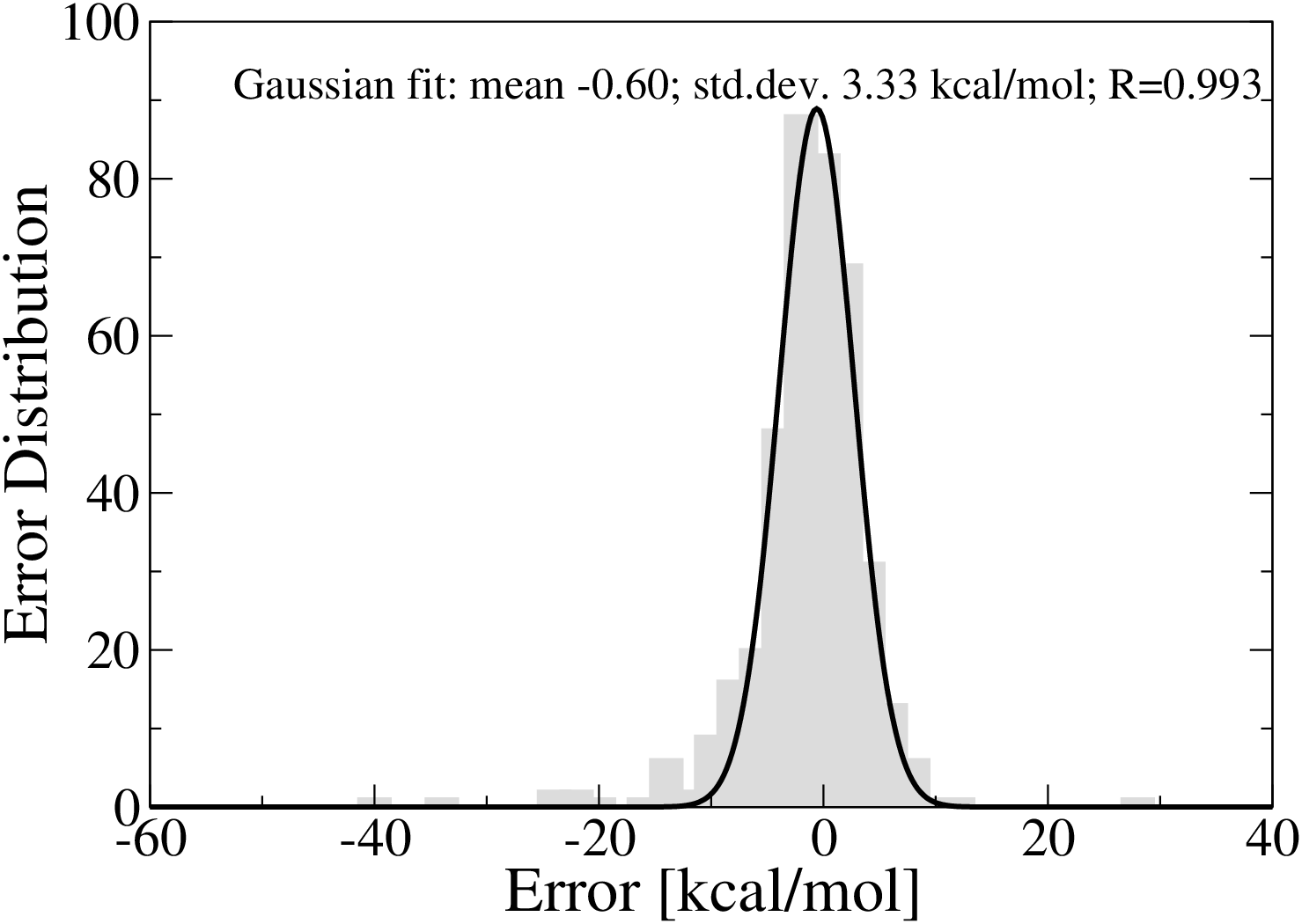}
\vspace{5cm}\\
%\caption{\label{FigS1}Error distribution of the BMK functional for %the large set of 402 energies
%excluding the reaction energies.}
\noindent {\rm Figure S-1: Error distribution of the BMK functional for the large set of 402 energies
excluding the reaction energies.}
\end{figure*}

\newpage
\clearpage
\begin{figure*}
\includegraphics[height=12cm,width=18cm]{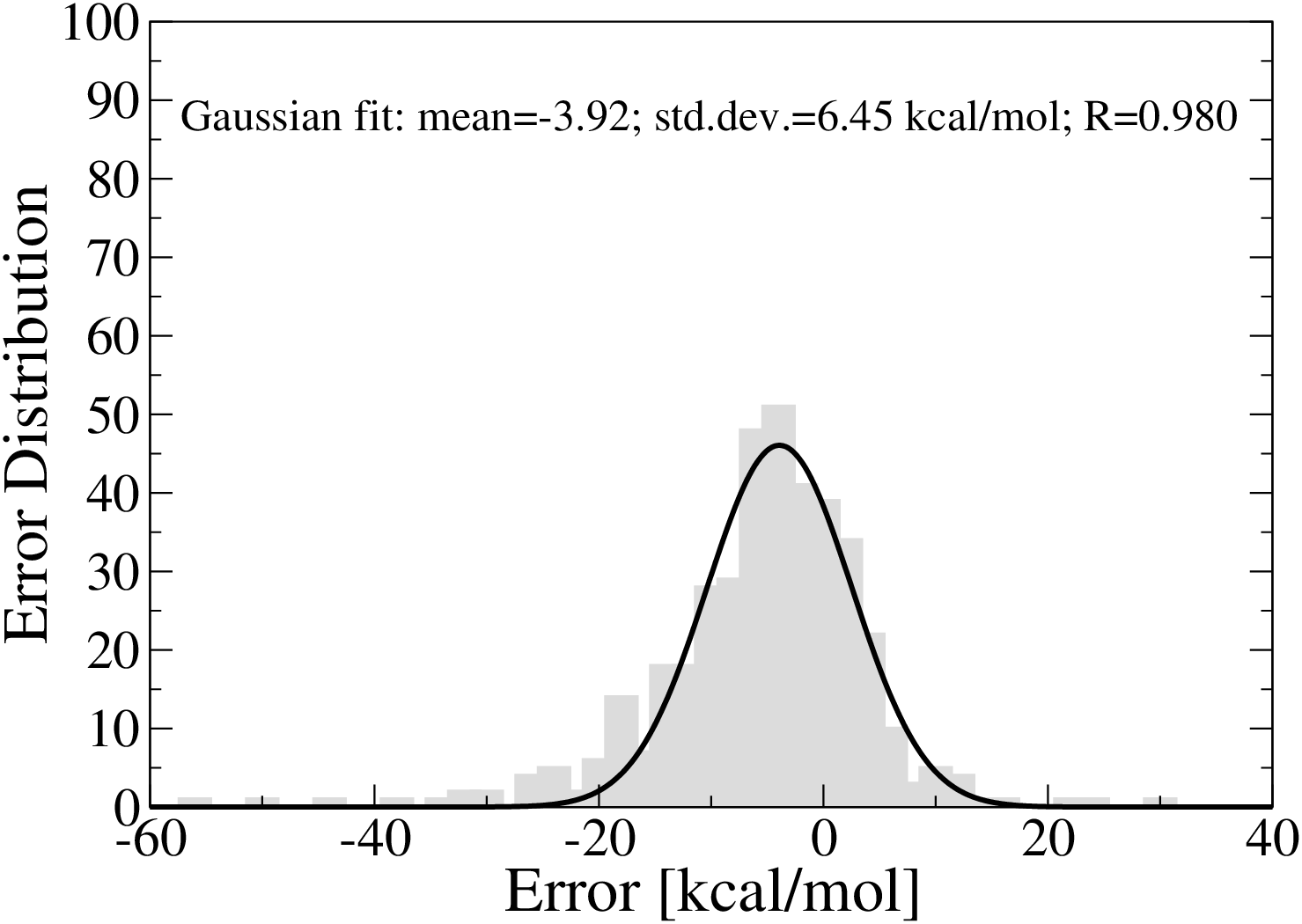}
\vspace{5cm}\\
\noindent {\rm Figure S-2: Error distribution of the B97-K functional for the large set of 402 energies
excluding the reaction energies.}
\end{figure*}

\newpage
\clearpage
\begin{figure*}
\includegraphics[height=12cm,width=18cm]{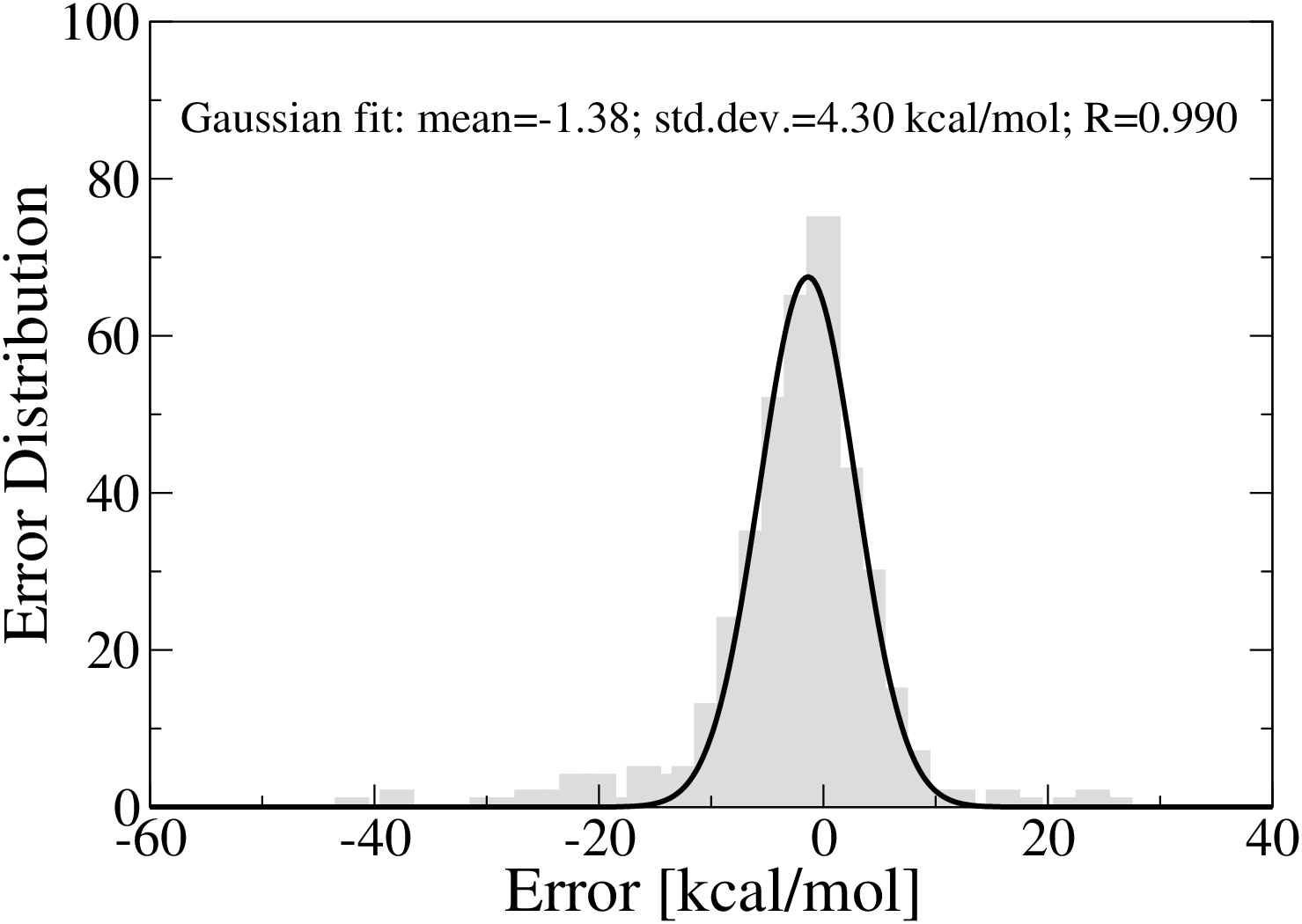}
\vspace{5cm}\\
\noindent {\rm Figure S-3: Error distribution of the B3LYP functional for the large set of 402 energies
excluding the reaction energies.}
\end{figure*}

\newpage
\clearpage
\begin{figure*}
\includegraphics[height=12cm,width=18cm]{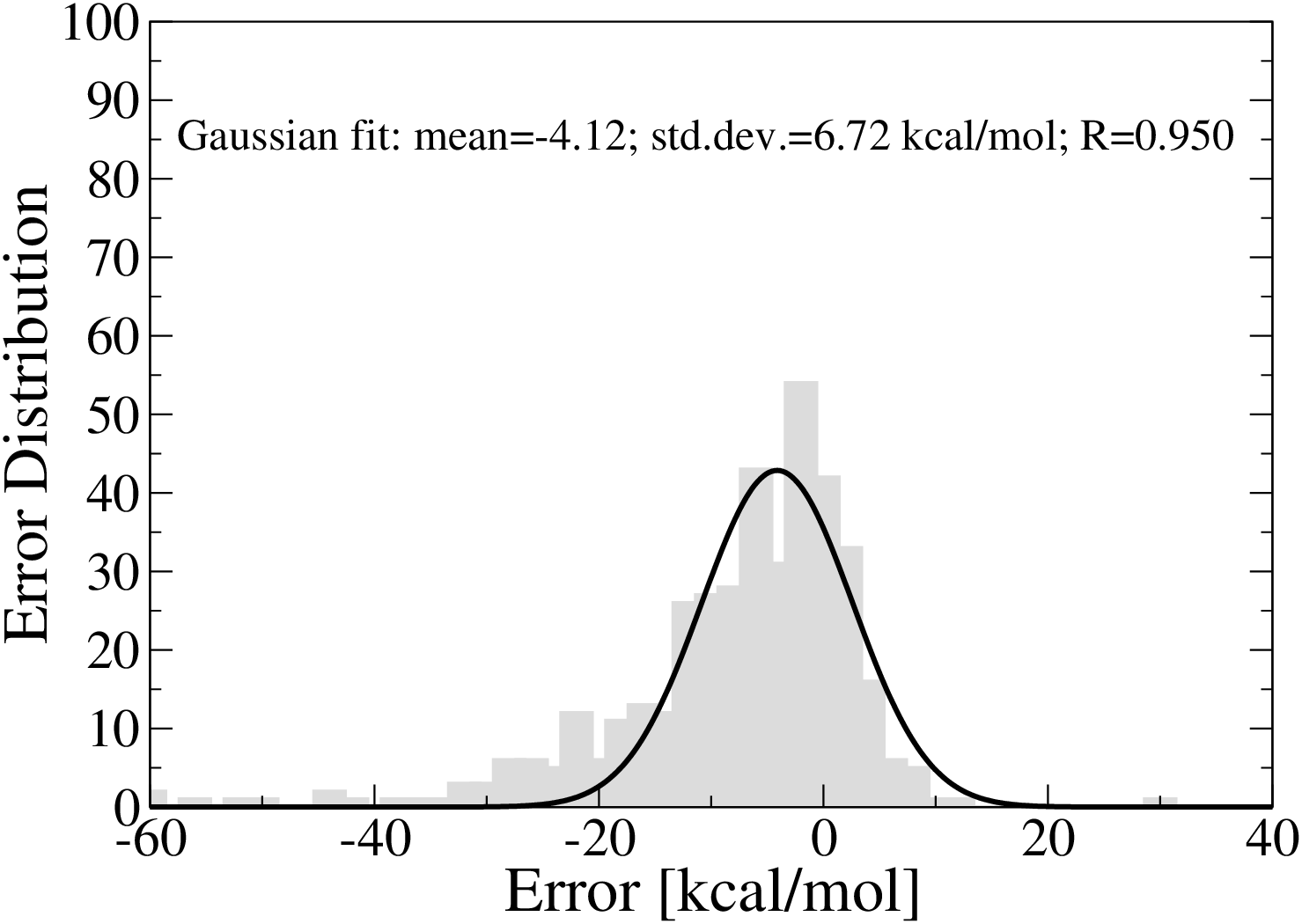}
\vspace{5cm}\\
\noindent {\rm Figure S-4: Error distribution of the mPW1K functional for the large set of 402 energies
excluding the reaction energies.}
\end{figure*}
\clearpage

\begingroup
\squeezetable
\begin{table*}[H]
\caption{Performance of new functionals for barrier heights (kcal/mol) of additional reactions.} 
\begin{tabular}{|l l l|l l l l|l l l l l l l|} \hline\hline
& &          &\multicolumn{4}{c|}{Reference
Method}&\multicolumn{7}{c|}{Functional Tested}\\\hline
\# & Set & Reaction
    & W1   & G2     & CBS-APNO/& exp.   & BMK  &
B97-K & mPW1K & KMLYP & B3LYP & B97-1 & B97-2\\ 
   &   &  
    & /W2   &       & /CBS-QB3 &    &    &
 &  &   & & & \\ \hline
 1 &   1 & C$_5$H$_8$          $\rightarrow$ C$_5$H$_8$
    & 36.8    &        &                  & 38.4   & 37.0 &
38.6 & 36.1   & 37.8  & 35.0 & 33.3 & 33.9 \\
 2 &   1 & C$_2$H$_6$ + NH     $\rightarrow$ C$_2$H$_5$ +
NH$_2$  & 19.4    &        &                  & 18.4   & 18.2
&  16.9 & 18.2 & 16.6 & 14.2 & 12.7 & 14.5 \\
 3 &   1 & C$_2$H$_6$ + NH$_2$ $\rightarrow$ C$_2$H$_5$ +
NH$_3$  & 11.3    &        &                  & 10.4   & 11.3
&  11.7 & 12.0 & 9.8  &  8.9 &  6.7 & 8.2 \\
 4 &   1 & C$_2$H$_6$ + OH     $\rightarrow$ C$_2$H$_5$ +
H$_2$O  &  3.5    &        &                  &  3.4   &  2.9
&   4.1 &  4.7 & 2.5  & -0.5 & -1.3 & 0.0 \\
 5 &   1 & CH$_3$     + OH     $\rightarrow$ CH$_4$     + O
    &  9.0    &        &                  &  7.8   &  6.3 &
6.7 &  7.6 &     5.4  &  4.6 &  2.3 & 3.1 \\
 6 &   1 & H$_2$      + Cl     $\rightarrow$ H          + HCl
    &  7.9    &        &                  &  8.7   &  5.1 &
5.3 &  4.7 &     2.9  &  5.2 &  3.9 & 5.2 \\
 7 &   1 & H$_2$      + F      $\rightarrow$ H          + HF
    &  2.3    &        &                  &  1.8   & -1.4 &
-0.7 &  0.5 &   -1.9  & -4.6 & -4.6 & -3.0 \\
 8 &   1 & CH$_3$     + HCl    $\rightarrow$ CH$_4$     + Cl
    &  1.4    &        &                  &  1.8   & -1.0 &
-0.8 &  0.5 &   -1.5  & -1.0 & -3.3 & -1.8 \\
 9 &   1 & H$_2$S     + H      $\rightarrow$ SH         +
H$_2$   &  3.6    &        &                  &  3.6   &  3.6
&   4.8 &  2.7 & 0.9  &  0.4 &  4.0 & 5.2 \\
10 &   1 & H$_2$      + H      $\rightarrow$ H          +
H$_2$   &  9.7    &        &                  &  9.6   &  9.5
&  10.2 &  7.1 & 4.9  &  4.1 &  8.8 & 9.8 \\
11 &   1 & CH$_3$OH   + H      $\rightarrow$ CH$_2$OH   +
H$_2$   &  9.6    &        &                  &  7.3   &  8.3
&   9.6 &  6.7 & 5.3  &  2.6 &  6.1 & 7.0 \\
12 &   1 & HCl        + H      $\rightarrow$ H          + HCl
    & 17.3    &        &                  & 18.0   & 17.2 &
19.3 & 16.3 & 15.7    & 12.7 & 16.5 & 18.3 \\
13 &   1 & N$_2$H$_2$ + H      $\rightarrow$ N$_2$H     +
H$_2$   &  2.8    &        &                  &  5.9   &  2.6
&   3.4 &  1.7 & -0.1 & -0.1$^*$ & 2.5 & 3.5 \\
14 &   1 & PH$_3$     + H      $\rightarrow$ PH$_2$     +
H$_2$   &  2.8    &        &                  &  3.2   &  3.8
&   4.2 &  2.1 &  0.3 &  0.2 &  3.6 & 4.6 \\
15 &   1 & OH         + H      $\rightarrow$ O          +
H$_2$   & 11.0    &        &                  & 10.1   & 10.3
&  12.3 &  9.1 &  7.4 &  4.1 &  7.8 & 8.3 \\
16 &   1 & OH         + Cl     $\rightarrow$ O          + HCl
    &         &        &                  &  9.8   &  7.9 &
10.0 &  9.0 &     8.1 &  5.3 &  2.9 & 3.4 \\
17 &   2 & CH$_3$Cl   + Cl$^-$ $\rightarrow$ Cl$^-$     +
CH$_3$Cl&  2.7    &        &                  &        &  3.6
&   1.8 &  3.6 &  3.0 & -0.7 & -1.0 & 1.6 \\
18 &   2 & CH$_3$F    + F$^-$  $\rightarrow$ F$^-$      +
CH$_3$F &  0.3    &        &                  &        &  1.8
&   0.3 &  2.0 &  0.6 & -2.3 & -2.2 & 0.2 \\
19 & 1,3 & CH$_3$     + H$_2$  $\rightarrow$ CH$_4$     + H
    & 11.9    &12.8$^b$&         11.3$^c$ &12.1,8.7$^a$& 10.0
&9.0 &  9.8&      7.44 & 8.7 &  7.2 & 8.4 \\
20 & 1,3 & H$_2$      + OH     $\rightarrow$ H          +
H$_2$O  &  5.4    & 6.8$^b$&          4.6$^c$ &5.7,3.6$^a$&
4.0 &4.7 &  4.8&  2.2 &  0.9 &  0.4 & 1.8 \\
21 & 1,3 & CH$_4$     + NH$_2$ $\rightarrow$ CH$_3$     +
NH$_3$  & 13.9    &15.0$^b$&         12.9$^c$ &15.5,12.6$^a$&
13.7&14.1& 14.4& 12.3 & 11.4 &  9.3 & 10.8 \\
22 & 1,3 & CH$_4$     + OH     $\rightarrow$ CH$_3$     +
H$_2$O  &  6.2    & 7.4$^b$&          4.9$^c$ &6.7,6.6$^a$&
5.6 &6.9 &  7.3&  5.2 &  2.2 &  1.1 & 2.5 \\
23 & 1,3 & NH$_3$     + OH     $\rightarrow$ NH$_2$     +
H$_2$O  &  3.5    & 5.1$^b$&          2.5$^c$ &3.2,1.9$^a$&
2.7 &4.1 &  5.0&  3.5 & -2.5 & -3.5 & -1.7 \\
24 & 1,3 & CH$_4$     + NH     $\rightarrow$ CH$_3$     +
NH$_2$  & 22.0    &23.3$^b$&         21.3$^c$ &22.7,20.5$^a$&
21.2&19.8& 21.2& 19.7 & 17.3 & 15.9 & 17.8 \\
25 &   4 & cyclobutene         $\rightarrow$ butadiene
    & 35.3$^d$&        &         33.7$^a$ &33.6$^a$& 40.4 &
37.4 & 39.8 &    41.4 & 33.0 & 34.6 & 35.8\\
26 &   4 & cis-1,3,5-hexatriene$\rightarrow$  &&&&&&&&&&&\\
 &  & ~~~cyclohexadiene
    &         &        &         28.8$^a$ &30.2$^a$& 31.9 &
33.2 & 31.3 &    32.7 & 31.2 & 29.4 & 30.1\\
27 &   4 & ortho-xylylene      $\rightarrow$ &&&&&&&&&&&\\
 & & ~~~benzocyclobutane
    &         &        &         25.9$^a$ &29.8$^a$& 29.2 &
30.2 & 28.5 &    30.8 & 28.3 & 26.4 & 26.7\\
28 &   4 & 1,3-pentadiene [1,5]-&&&&&&&&&&&\\
   &    &  ~~~sigmatropic H-shift
    &  39.5$^e$ &        &         38.9$^a$ &38.8$^a$& 40.3 &
42.1 & 39.7 &    41.2 & 38.9 & 36.9 & 37.6\\
29 &   4 & 1,3-cyclopentadiene [1,5]-&&&&&&&&&&&\\
   &     &  ~~~sigmatropic H shift  &  28.2$^e$&    & 28.1$^a$ &26.0$^a$& 27.5 & 
28.8 & 26.2 &    27.0 & 27.2 & 26.4 & 25.3\\
30 &   4 & 1,5-hexadiene [1,5]-&&&&&&&&&&&\\
   &     &  ~~~sigmatropic H-shift
    &         &        &         33.3$^a$ &34.8$^a$& 36.7 &
39.1 & 37.9 &    40.2 & 26.4 & 33.6 & 35.3\\
31 &   4 &ethylene + butadiene $\rightarrow$ &&&&&&&&&&&\\
  &   &  ~~~cyclohexene
    &         &        &         20.7$^a$ &21.1$^a$& 25.4 &
27.2 & 23.8 &    21.9 & 26.4 & 22.0 & 24.6\\
32 &   4 &ethylene + cyclopentadiene $\rightarrow$&&&&&&&&&&&\\
   &     & ~~~ norbonene
    &         &        &         15.1$^a$ &19.4$^a$& 21.8 &
23.8 & 20.3 &    18.2 & 24.4 & 19.4 & 22.0\\
33 &   4 & 2 cyclopentadiene
$\rightarrow$&&&&&&&&&&&\\
   &     &  ~~~endo-dicyclopentadiene&         &        &
   9.8$^a$ &13.3$^a$& 20.1$^f$ &  23.2$^f$ & 19.7$^f$ & 21.7$^f$ & 23.9$^f$ & 18.4$^f$ & 21.7\\
34$^f$ &   4
&cis-triscyclopropacyclohexane$\rightarrow$&&&&&&&&&&&\\
   &    
& ~~~cyclonona-1,4,7-triene&
&        &         23.5$^a$ &26.4$^a$& 22.1$^f$ & 25.5$^f$ & 33.3$^f$ &
36.6$^f$ & 21.9$^f$ & 20.4$^f$ & 27.4$^f$\\
35 &  
&cis-triscyclobutacyclohexane$\rightarrow$&&&&&&&&&&&\\
   &    
& ~~~cyclodeca-1,5,9-triene&
 &        &                  &50.0$^a$& 67.7$^f$ & 60.3$^f$ & 65.5$^f$ & 70.3$^f$
& 48.8$^f$ & 52.8$^f$ & 55.1\\
\hline
\end{tabular}
(continued on next page)
\end{table*}
\endgroup
\pagestyle{empty}
\clearpage
\addtocounter{table}{-1}
\begingroup
\squeezetable
\begin{table*}[H]
\caption{(continued)}
\begin{tabular}{|l l l|l l l l|l l l l l l l|} \hline
& &          &\multicolumn{4}{c|}{Reference
Method}&\multicolumn{7}{c|}{Functional Tested}\\\hline
\# & Set & Reaction
    & W1   & G2     & CBS-APNO/& exp.   & BMK  &
B97-K & mPW1K & KMLYP & B3LYP & B97-1 &B97-2 \\ 
   &   &  
    & /W2   &       & /CBS-QB3 &    &    &
 &  &   & & & \\ \hline
36 &   3 & C$_2$H$_6$ + H      $\rightarrow$ C$_2$H$_5$ +
H$_2$   &         &13.4$^b$&          9.4$^c$ & 8.8$^a$& 11.4
&  13.5 & 10.5 &  9.0 &  6.6 & 10.2 & 11.0  \\
37 &   3 & CH$_3$     + CH$_4$ $\rightarrow$ CH$_4$     +
CH$_3$  &         &18.4$^b$&         16.6$^c$ &15.1$^a$& 17.0
&  16.1 & 17.4 & 15.5 & 15.7 & 13.1 & 14.5  \\
38 &   3 & CH$_3$F    + H      $\rightarrow$ CH$_2$F    +
H$_2$   &         &14.2$^b$&         12.9$^c$ &10.4$^a$& 13.3
&  14.6 & 11.5 & 10.2 &  7.5 & 10.9 & 11.8  \\
39 &   3 & NH$_3$     + H      $\rightarrow$ NH$_2$     +
H$_2$   &         &16.3$^b$&         14.9$^c$ &10.1$^a$& 15.7
&  15.9 & 14.1 & 13.2 &  9.0 & 12.7 & 13.9 \\
40 &   3 & NH$_3$     + O      $\rightarrow$ NH$_2$     + OH
    &         &16.7$^b$&         12.2$^c$ &11.4$^a$& 10.6 &
9.8 & 12.5 &     11.8 &  4.3 &  4.0 & 6.5 \\
41 &   3 & CH$_4$     + O$_2$  $\rightarrow$ CH$_3$     +
HO$_2$  &         &55.6$^b$&         56.7$^c$ &56.0$^c$& 56.2
&  55.0 & 55.2 & 54.4 & 52.8 & 53.3 & 56.6 \\
42 &   3 & CH$_3$F    + O      $\rightarrow$ CH$_2$F    + OH
    &         &14.3$^b$&         11.7$^c$ &11.0$^a$&  9.4 &
9.8 & 11.6 &     10.3 &  4.8 &  3.9 & 5.8 \\
43 &   3 & CH$_3$F    + OH     $\rightarrow$ CH$_2$F    +
H$_2$O  &         & 6.7$^b$&                  & 4.2$^a$&  3.3
&   5.0 &  5.3 &  3.2 & -0.0 & -1.1 & 0.3 \\
44 &   3 & CH$_2$F$_2$+ O      $\rightarrow$ CHF$_2$    + OH
    &         &15.2$^b$&                  &11.4$^a$&  9.9 &
10.4 & 12.1 &    11.0 &  5.0 &  4.0 & 5.9 \\
45 &   3 & CHF$_3$    + OH     $\rightarrow$ CF$_3$     +
H$_2$O  &         & 7.5$^b$&                  & 7.8$^a$&  6.7
&   8.4 &  8.6 &  6.8 &  2.9 &  1.5 & 2.8 \\
46 &   3 & CH$_3$Br   + O      $\rightarrow$ CH$_2$Br   + OH
    &         &14.4$^b$&                  &12.1$^a$& 10.2 &
10.4 & 12.5 &    11.2 &  5.2 &  4.4 & 6.3 \\
47 &   3 & CHF$_3$    + O      $\rightarrow$ CF$_3$     + OH
    &         &23.9$^b$&                  &15.9$^a$& 14.6 &
14.7 & 16.6 &    15.7 &  9.4 &  8.3 & 10.2 \\
48 &   3 & CH$_2$F$_2$+ OH     $\rightarrow$ CHF$_2$    +
H$_2$O  &         & 6.1$^b$&                  & 4.8$^a$&  3.4
&   5.1 &  5.3 &  3.4 & -0.1 & -1.2 & 0.1 \\
49 &   3 & CF$_3$     + H$_2$  $\rightarrow$ CHF$_3$    + H
    &         &10.9$^b$&                  & 9.4$^a$& 11.0 &
9.6 & 10.3 &      7.6 &  9.4 &  8.5 & 9.9 \\
50 &   3 & CH$_3$Br   + OH     $\rightarrow$ CH$_2$Br   +
H$_2$O  &         & 6.2$^b$&                  & 5.1$^a$&  4.0
&   5.6 &  6.0 &  3.8 &  0.5 & -0.6 & 0.8 \\
51 &   3 & CH$_2$Br   + H$_2$  $\rightarrow$ CH$_3$Br   + H
    &         &14.2$^b$&                  &11.5$^a$& 13.1 &
11.8 & 12.6 &    10.4 & 11.9 & 10.7 & 12.3 \\
52 &   3 & C$_2$H$_6$ + CH$_3$ $\rightarrow$ C$_2$H$_5$ +
CH$_4$  &         &16.0$^b$&                  &12.8$^a$& 15.1
&  14.3 & 15.4 & 13.4 & 13.7 & 11.0 & 12.3 \\
53 &   3 & SiH$_4$    + H      $\rightarrow$ SiH$_3$    +
H$_2$   &         & 5.6$^b$&                  & 3.0$^a$&  5.6
&   6.5 &  3.6 &  1.9 &  1.3 &  5.1 & 5.8 \\
54 &   3 & NH$_3$     + NH$_2$ $\rightarrow$ NH$_2$     +
NH$_3$  &         &14.6$^b$&         13.7$^c$ &        & 10.9
&  10.9 & 12.0 & 10.5 &  7.0 &  5.2 & 7.0 \\
55 &   3 & PPH$_3^+$           $\rightarrow$ HP-PH$_2^+$
    &         &39.3$^b$&                  &        & 41.9 &
41.3 & 41.2 &    42.4 & 39.2 & 38.9 & 37.9 \\
56 &   3 & CH$_4$     + NO$_2$ $\rightarrow$ CH$_3$     +
HNO$_2$ &         &35.9$^b$&                  &        & 36.9
&  36.7 & 34.5 & 32.5 & 34.3 & 33.3 & 35.1 \\
57 &   3 & H$_3$CO             $\rightarrow$ H$_2$COH
    &         &32.9$^b$&         31.4$^c$ &        & 34.5 &
36.8 & 35.6 &    36.2 & 33.5 & 32.9 & 32.6 \\
58 &   3 & HO$_2$     + NH     $\rightarrow$ O$_2$      +
NH$_2$  &         & 1.9$^b$&          0.6$^c$ &        & -0.0
&   0.2 &  1.2 &  0.1 & -4.4 & -5.5 & -4.0 \\
59 &   3 & CH$_2$F$_2$+ NH     $\rightarrow$ CHF$_2$    +
NH$_2$  &         &21.7$^b$&         13.0$^c$ &        & 18.0
&  17.2 & 18.5 & 17.2 & 13.9 & 12.2 & 14.0 \\
60 &   3 & NH$_2$     + O      $\rightarrow$ NH         + OH
    &         & 9.9$^b$&          6.3$^c$ &        &  4.7 &
5.8 &  7.3 &      5.8 & -0.5 & -1.4 & 0.1 \\
61 &   3 & CH$_3$     + O      $\rightarrow$ CH$_2$     + OH
    &         &17.0$^b$&         14.9$^c$ &        & 13.4 &
15.6 & 15.4 &    14.0 & 10.0 &  8.6 & 10.2 \\
62 &   3 & CH$_3$CH$_2$        $\rightarrow$ CH$_2$CH$_3$
    &         &42.8$^b$&         42.2$^c$ &        & 42.6 &
42.3 & 43.0 &    43.1 & 41.5 & 40.7 & 40.3 \\
63 &   3 & CH$_3$F    + H      $\rightarrow$ CH$_3$     + HF
    &         &31.4$^b$&         30.9$^c$ &30.6$^a$& 31.4 &
31.1 & 30.5 &    30.4 & 21.7 & 27.4 & 29.5 \\
64 &   3 & CF$_4$     + H      $\rightarrow$ CF$_3$     + HF
    &         &40.9$^b$&                  &44.3$^a$& 44.2 &
44.2 & 44.2 &    46.0 & 32.7 & 38.5 & 40.7 \\
65 &   3 & CH$_3$Br   + H      $\rightarrow$ CH$_3$     + HBr
    &         & 6.1$^b$&                  & 5.9$^c$&  6.5 &
7.7 &  5.1 &      3.4 &  2.0 &  5.7 & 7.1 \\
66 &   3 & SO         + O$_2$  $\rightarrow$ SO$_2$     + O
    &         &18.3$^b$&                  & 4.6$^a$&  6.4 &
5.0 & 10.3 &      5.7 &  6.3 &  5.0 & 8.4 \\
67 &   3 & CO         + O$_2$  $\rightarrow$ CO$_2$     + O
    &         &52.3$^b$&                  &47.2$^a$& 51.8 &
51.7 & 48.1 &    47.4 & 47.2 & 47.1 & 49.7 \\
68 &   3 & SCO        + O      $\rightarrow$ SO         + CO
    &         & 2.4$^b$&                  & 5.4$^a$&  3.0 &
2.9 &  5.9 &      3.8 & -1.6 & -2.5 & -0.4 \\
69 &   3 & CO         + BrO    $\rightarrow$ CO$_2$     + Br
    &         &11.4$^b$&                  & 7.4$^a$& 12.8 &
13.3 & 12.4 &    10.5 &  9.6 &  8.6 & 10.2 \\
70 &   3 & N$_2$      + H      $\rightarrow$ N$_2$H
    &         &14.5$^b$&         14.1$^c$ &        & 13.3 &
14.6 & 11.2 &     9.5 &  8.2 & 12.0 & 13.2 \\
71 &   3 & N$_2$O     + Br     $\rightarrow$ N$_2$O     + BrO
    &         &        &                  &33.0$^a$& 31.8 &
31.0 & 35.3 &    32.2 & 30.4 & 29.1 & 34.2 \\
\hline\hline\end{tabular}

\vspace{12pt} \noindent $^a$ Energy has been corrected by scaled
(0.9804) B3LYP/6-31G* ZPVE to obtain bottom-of the well
values.\\
\noindent $^b$ Energy has been corrected by scaled
(0.8929) HF/6-31G* ZPVE to obtain bottom-of the well
values.\\
\noindent $^c$ Energy has been corrected by scaled
(0.92511) HF/6-311G** ZPVE to obtain bottom-of the well
values.\\
\noindent $^d$ At mPW1K/6-31+G* geometry.\\
\noindent $^e$ Reactant at B3LYP/6-31+G* geometry, transition state at mPW1K/6-31+G* geometry.\\
\noindent $^f$ Using a 6-311+G(2d,p) basis set to optimize the geometries.\\
\noindent $^*$ Dissociated to the products.\\
\end{table*}
\endgroup

\clearpage


\begin{thebibliography}{99}
\bibitem{forty-two} Any similarity to the ``ultimate answer'' obtained in the late Douglas Adams' classic series ``The Hitch-hiker's Guide to the Galaxy'' is purely coincidental.
\end{thebibliography}
\end{document}